\documentclass[onecolumn, draftcls, 11pt]{IEEEtran}
\pdfoutput=1
\usepackage{epsfig,latexsym,amssymb,amsmath,graphics}
\usepackage{graphicx}
\usepackage{amssymb,dsfont,amsthm}
\usepackage{cite,color}
\usepackage{url}
\usepackage{booktabs}
\usepackage{algorithm, algorithmicx}
\usepackage{algpseudocode}
\usepackage{stfloats}
\usepackage{verbatim}
\usepackage{epstopdf}
\usepackage{subfig}
\usepackage{bm}

\definecolor{ColorRed}{rgb}{1,0,0}

\allowdisplaybreaks[4]

\newtheorem{theorem}{Theorem}

\begin{document}
\title{The design and optimization of synchronization sequence  for Ultraviolet communication}

\author{Shihui Yu, Chen Gong, and Zhengyuan Xu
%
%
}
\maketitle

\begin{abstract}
	
In the ultraviolet (UV) scattering communication, the received signals exhibit the characteristics of discrete photoelectrons due to path loss. The synchronization is based on maximum Pulse Number-Sequence correlation problem. First of all, the accuracy of synchronization is vital to channel estimation and decoding. This article focuses on improving synchronization accuracy by designing and optimizing synchronization sequences. As for the maximum Pulse Number-Sequence correlation problem, it is assumed that the correlation values satisfy the Gaussian distribution and their mathematical expectation, variance and covariance are derived to express the upper bound of synchronization offset. The synchronization sequence we designed has two equilong RANDOM parts (Symbols meet Bernoulli distribution with equal probability.) and a $\{1,0,1,0,1,0,...,1,0,1,0\}$ part between them with $ \alpha $ as its proportion of entire sequence. On the premise of ensuring the synchronization reliability, the synchronization deviation can be reduced by optimizing $ \alpha $. 

There are simulation experiments to verify correctness of the derivation, reasonableness of the hypothesis and reliability of optimization. Compared with equilong random sequence, the synchronization accuracy of the optimized synchronization sequence is significantly improved.

\end{abstract}


\section{Introductions}

Ultraviolet communication with ultraviolet light as a communication medium is a kind of wireless optical communication, which has been widely concerned. Considering the absorption of ultraviolet light by atmosphere, solar blind ultraviolet light in UV-C band is introduced in the 1960s. Compared with the initial laser/flashtube/lamp as the emission source, the current UV communication system can apply better optical materials and hardware equipments, such as low cost semiconductor laser diodes and miniaturized light emitting diodes (LEDs) at UV-C frequencies \cite{8641355,10.1117/12.500861}, which makes UV communication exhibit some notable characteristics. First, in-band background noise close to the earth surface is almost negligible because it is solar blind. Hence, the transmission channel is almost noiseless and ideal\cite{1527982}. Second, ultraviolet has stronger aerosol and molecular scattering, and can transmit signals in a non-direct way. Strong scattering makes it possible that we can operate highly sensitive wide FOV quantum noise-limited photon counting Rxs \cite{10.1117/12.582002}. Third, UV-C channels are quite robust to meteorological conditions. However, channel attenuation in UV communication is a challenge, which reduces the maximum achievable data rate and the transmission range. Thus, dense network configurations as well as more sensitive sensing and signal recognition are important.

In the ultraviolet (UV) scattering communication, continuous-time and discrete-time Poisson channel' capacity have been explored in \cite{75239,21284,21285,6685986,6685979,4729780}. Besides, the UV photon-counting has been studied in \cite{8552373}. In \cite{5599260}, the generalized maximumlikelihood sequence detection has been researched. For the experimental evaluation of NLOS UV-C links outdoor, \cite{10.1117/12.735183} has conducted a communication test-bed.  the researchers at MIT Lincoln laboratory have done various outdoor experiments for a short range UV-C link \cite{10.1117/12.500861,10.1117/12.407519,10.1117/12.438327,10.1117/12.666194}. Researchers from the University of Virginia employed M-array amplification on detection \cite{6461162}. Experient work on long-distance channel characterization is introduced in \cite{6923957}. \cite{7112175,7047703} studied the signal detection with receiver diversity. \cite{5379755} exploited the solar blind UV spectrum to provide a short range, medium bandwidth, NLOS, networked communications system as an alternative to traditional RF communications systems. In \cite{8332484}, researchers have designed the system and finished hardware realization based on receiver diversity for the NLOS UV scattering communication over 1 km, where the system throughput can reach 1 Mbps. Frame synchronization employed in communication systems is to find valid data in a transmission by inserting a fixed data pattern. Binary sequence with perfect auto-correlation has been studied in \cite{Arasu2011SequencesAA}. There are periodical sequences' detection in a bit stream for synchronization \cite{1094813}. Researchers proposed a extension of joint frame synchronization of MIMO-OFDM systems \cite{4401715}. Rather than traditional synchronization techniques that employed the “same” fixed pattern, perfect punctured binary sequence pairs \cite{7042838} are applied to a new synchronization scheme \cite{article}. Mismatched filtering is introduced in \cite{1214826}, where the transmitter and the sender can use different sequences as a sequence pair. Such pairs attracted a lot of researchers owing to correlation properties.

High speed communication over long distances \cite{8332484} challenges the reliability and accuracy of synchronization processes. Mostly, m-sequence is used as the synchronization sequence in NLOS UV scattering communication systems. Locating the starting moment of frame header successfully benefits accurate channel estimation and decoding. Most researches about frame synchronization is just to locate the first symbol as the frame header. So estimation offset within the duration of a symbol can still exist, which may even cause the transmission failure when the signal - noise condition is poor.

Synchronization relies on the correlation of pulse numbers and the pretreated sequence \cite{8332484}. Our work combined with UV scattering communication systems aim to locate the moment when the Poisson process of photons'arriving corresponding to the first symbol starts more precisely. We design the structure of the sequence itself to explore the possibility of improving synchronization accuracy by decreasing estimation offset of the starting moment, even if it is within the duration of a symbol. This gives an indication to select synchronization sequences in a specific scenario of UV scattering communication.

 In this paper, we provide the Poisson channel model and the synchronization process in Section \ref{sec.system_model}. In Section \ref{section.Sequence Design}, the structure of synchronous sequence is designed based on maximum Pulse Number-Sequence correlation problem. In Section \ref{section.Mathematical expectation,variance and covariance of correlation values}, statistical analysis of correlation values is introduced for the synchronization with sequence we designed. In Section \ref{section.estimation of the synchronization accuracy}, it is assumed that the correlation values satisfy the Gaussian distribution. Therefore, the estimating biases of synchronization process are quantified through calculation of the upper bound of the probability function. The optimization schemes of synchronous sequences are proposed in Section \ref{section.optimization of the synchronization sequence}. For vertification, simulation results  are shown in Section \ref{section.verification by simulating}. Finally, we conclude this paper in Section \ref{section.conclusion}.

\section{System model} \label{sec.system_model}
\subsection{Channel Model}
Consider UV scattering communication system with on-off keying (OOK) modulation. Due to extremely weak received signal can be characterized by discrete photoelectrons, whose arrival satisfies Poisson process. In particular, the number of photons received during a certain duration yields Poisson distribution. Let $\lambda_s$ and $\lambda_b$ denote the mean numbers of detected photoelectrons for the signal and background radiation components, respectively. Let $N_i$ denote the number of received photoelectrons from the $i_{th}$ symbol, denoted as $s_i \in \left\{0,1\right\}$, satisfying
\begin{align} \label{equ.Poisson_channel}
	P(N_i = n_i \mid s_i=1) &= \frac{(\lambda_s+\lambda_b)^{n_{i}}}{{n_{i}}!}e^{-(\lambda_s+\lambda_b)},\\
	P(N_i = n_i \mid s_i=0) &= \frac{\lambda_b^{n_{i}}}{{n_{i}}!}e^{-\lambda_b},
\end{align} 


\subsection{Synchronization Process}
Let $\{s_{i}:1\leq i \leq L\}$ denote the synchronization sequence, where $L$ is the length. We adopt pulse-counting-based digital detection. Let $ t_S $ denote the symbol duration. In this work, the received signal is divided into chips with duration $T_c = \frac{T_s}{n}$. For a certain chip index $t$, the number of pulses in that chip is denoted as $ C_{t} $. Let $\widehat{t_{start}}$ denote the estimated starting chip index of the synchronization sequence. The synchronization solution is formulated by the following problem,
\begin{equation}
	\begin{aligned} \label{equ.auto-correlation}
		\widehat{t_{start}}&=\arg \max _{t_{start}}\left\{\sum_{i=1}^{L}\left\{n_{i} \times\left(2 s_{i}-1\right)\right\}\right\},\\
		n_{i}&=\sum_{j=0}^{n-1}C_{t_{start}+n\times (i-1) +j}~,
	\end{aligned}
\end{equation}
 where $ n_{i} $ denotes the numbers of pulses correlated with symbol $ s_{i} $ based on $ t_{start} $, and symbol $ s_{i} $ is transformed to $ \left(2 s_{i}-1\right) $ belonging to $ \{-1,1\} $ for correlation. The maximum correlation peak criterion is adopted.

\section{the synchronization sequence structure} \label{section.Sequence Design}
%
In the chip-level synchronization approach, the received symbol is divided into chips and the most accurate starting chip is estimated via maximum correlation peak criterion as shown in Eq.\eqref{equ.auto-correlation}, where the synchronization performance is related to the synchronization sequence structure. In this Section, we propose ``Sandwich" structure of the synchronization sequence, with a parameter on the ratio of each component to be optimized. 
\subsection{Structure of Synchronization Sequence}
\begin{figure}[htbp]
	\centering
	\includegraphics[width=5.0in]{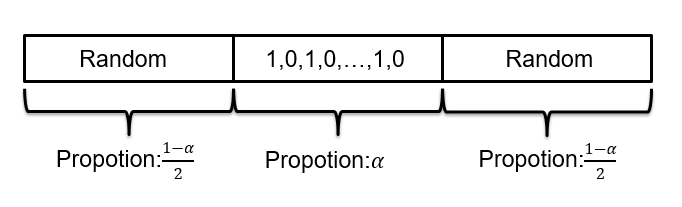}
	\caption{The structure of the synchronization sequence.} \label{SeqStruc}
\end{figure}

The ``Sandwich" structure of the sequence is shown in Fig \ref{SeqStruc}. We adopt alternate sequence $\{1,0,1,0,1,0,...,1,0,1,0\}$ as the middle part of the synchronization sequence, whose propotion is denoted as $ \alpha $. Both sides of the synchronization sequence consists of two equal-long and $0-1$ equal-probablility RANDOM parts, each with proportion $ \frac{1-\alpha}{2} $. 

\subsection{Synchronization Misestimate Events}\label{misesti}

\begin{figure}[htbp]
	\centering
	\includegraphics[width=3.5in]{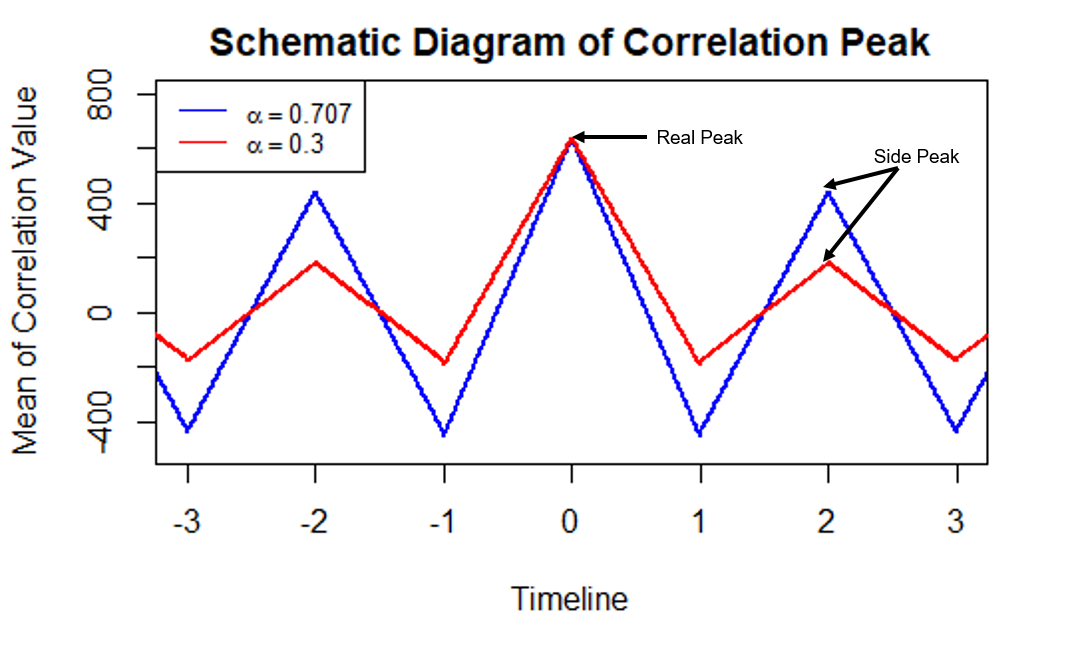}
	\caption{The schematic diagram of correlation value under different $ \alpha $.} \label{peak}
\end{figure}

Consider the offset within the duration of a symbol. In such case, the number of photons that should be correlated with $s_i$ is partially correlated with its neighbor $s_{i+1}$ or $s_{i-1}$. If $s_i$ is the same as its neighbor $s_{i+1}$ or $s_{i-1}$, such offset does not affect the correlation value. On the contrary, the correlation will change when $s_i$ is different from its neighbor $s_{i+1}$ or $s_{i-1}$, which implies that the correlation is sensitive to $\widehat{t_{start}}$ if more symbols are different from their neighbors. Sequence $\{1,0,1,0,1,0,...,1,0,1,0\}$ is adopted since each symbol has two different neighborhoods. Blue line in Fig. \ref{peak} indicates that the real peak of correlation value is sharper when $ \alpha $ is relatively large, which benefits the location of correlation peak. We obtain two groups of results in Fig. \ref{peak} from theoritical calculation introduced in Section. \ref{section.Mathematical expectation,variance and covariance of correlation values}. Other parameters are the same except $ \alpha $ in two groups, such as $ L=128,\lambda_s=10,\lambda_{b}=1,n=100 $.

On the other hand, $0$-$1$ alternate sequence $\{1,0,1,0,1,0,...,1,0,1,0\}$ will produce non-negligible correlation peak at the offset of even symbols, for example $2$ and $4$ symbols. Fig \ref{peak} shows that the peak at offset $2$ symbols becomes higher for larger $ \alpha $, which distracts location of the real peak. Consider an extreme case where the synchronization sequence consists of $\{1,0,1,0,1,0,...,1,0,1,0\}$. The correlation value rouphly decreases until the offset reaches a symbol at a local lowest point, and then keeps increasing to a side peak value at offset of two symbols. These side peaks at offset of even symbols from the main peak may lead significant loss of synchronization accuracy. Hence, part of the synchronization sequence need to be random to suppress the peak at offset of even symbols. Proportion $ \alpha $ needs to be optimized.

\section{Statistical Properties of correlation values}  \label{section.Mathematical expectation,variance and covariance of correlation values}
We aim to analyze the MSE of synchronization error based on correlation peak criterion. To begin with, we analyze the statistical properties on the correlation values on potential peaks up to the second-order moments.  
\subsection{Correlation Value Calculation}
\begin{figure}[htpb]
	\centering
	\includegraphics[width=6.5in]{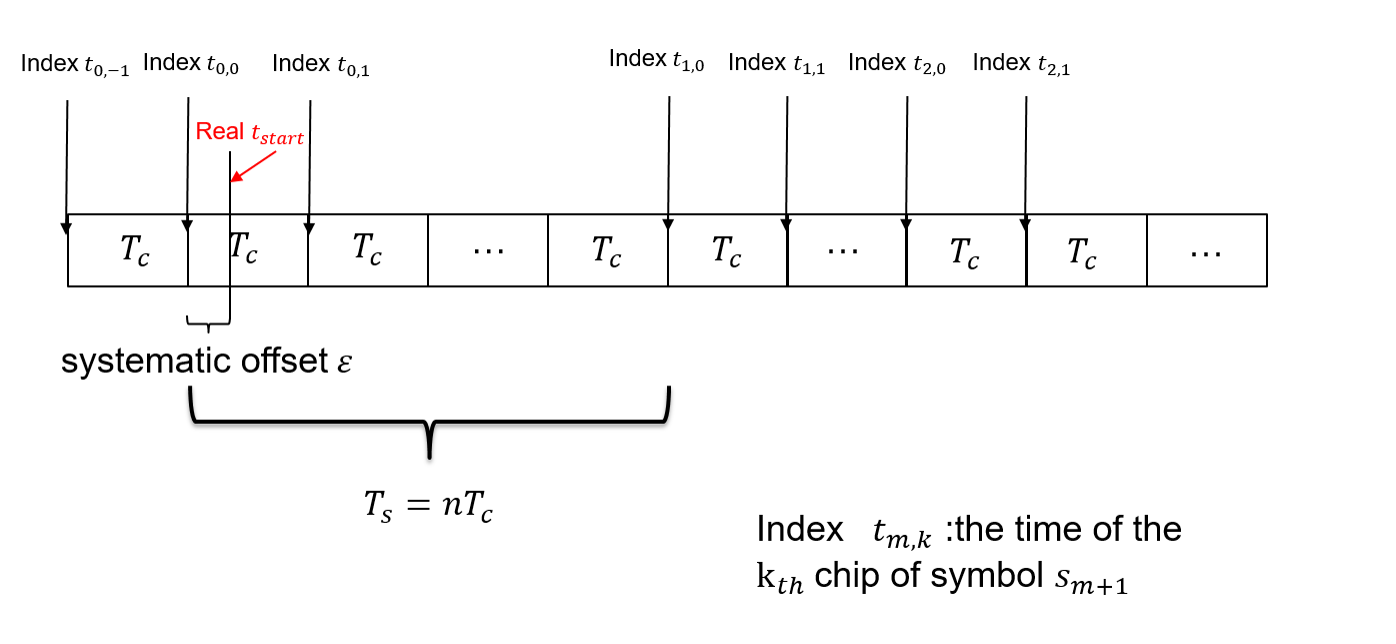}
	\caption{The timeline of the synchronization.} \label{timeline}
\end{figure}

The synchronization timeline is visualized in Fig \ref{timeline}.

The process of dividing time into chips introduces systematic offsets, denoted as $ \epsilon $ satisfying uniform distribution $ U(-\frac{T_c}{2},\frac{T_c}{2}) $. Assuming $t_{0,,0} = 0$ as the closest and most accurate timeIndex. Let estimate $\widehat{t_{start}}=t_{0,0} $ denote the accurate synchronization estimate as Case 1. The correlation value is denoted as $ C_{m,k} $ when we treat $ t_{m,k} $ as the starting timeIndex.

Synchronization offset happens if $C_{0,0}$ is not the largest correlation, which can be summarized into other two cases. In Case 2, the offset within a symbol's duration,i.e., $\widehat{t_{start}}=t_{0,k} $ for certain $ 1\leq \vert k \vert \leq n-1 $. To cope with the problem that $ C_{0,k} \geq C_{0,0}$, the expectation and variance of $ C_{0,k} $, as well as the covariance between $C_{0,k}$ and $ C_{0,0} $, are analyzed. In Case 3, consider the offset over a symbol duration,  $\widehat{t_{start}}=t_{2m,k} $ where where $ \vert m \vert \geq 1  $ and $ -n \leq  k  \leq n-1 $. Similarly, the expectation and variance of $ C_{2m,k} $, as well as the covariance between $C_{2m,k}$ and $ C_{0,0} $, are analyzed, to cope with the problem that $ C_{2m,k} \geq C_{0,0}$. 

Giving a fixed synchronization sequence and systematic offset $ \epsilon $, we can get the conditional expectation, variance and covariance based on the Poisson arrival process.

\subsection{Case 1 : Estimate $\widehat{t_{start}}$ is chip $t_{0,0}$}
 As the arriving of photons is the Poisson process, we can calculate $ C_{0,0} $ as follows,

\begin{equation} \label{equ.correlation for case 1}
\operatorname{C}_{0,0}  = \sum_{i=1}^{L}\left\{n_{i} \times\left(2 s_{i}-1\right)\right\}
\end{equation}

\begin{equation}\label{pnum1}
n_{i} \sim \left[\operatorname{Poi}\left(\varepsilon\left(\lambda_{s} s_{i-1}+\lambda_{b}\right)\right)+\operatorname{Poi}\left((1-\varepsilon)\left(\lambda_{s} s_{i}+\lambda_{b}\right)\right)\right]
\end{equation} 

\begin{figure}[htpb]
	\centering
	\includegraphics[width=6.5in]{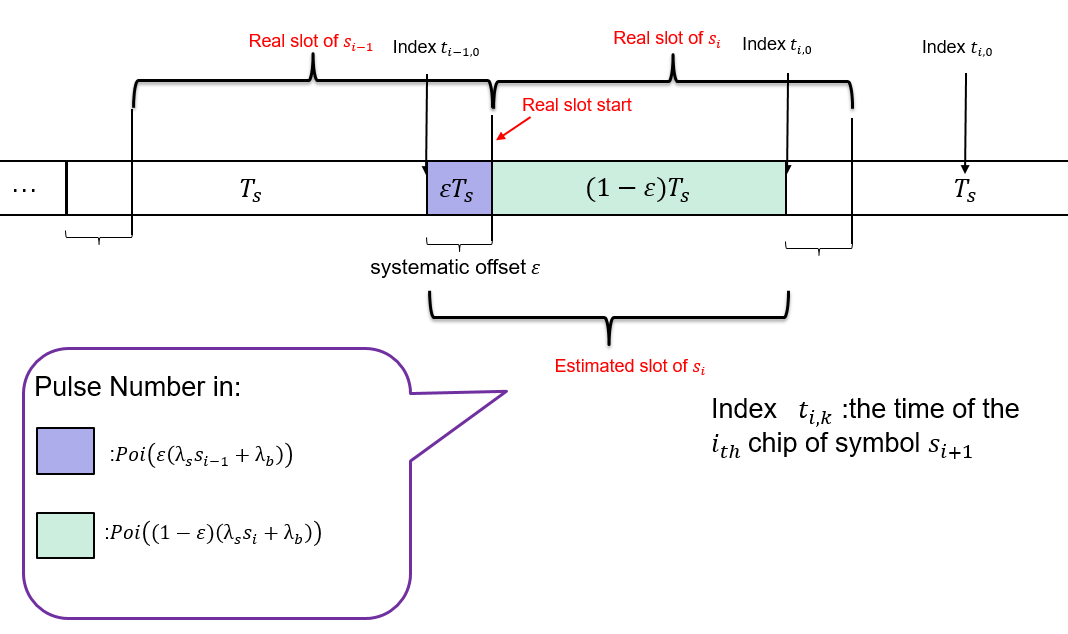}
	\caption{The timeline of the synchronization.} \label{timeline2}
\end{figure}
Fig. \ref{timeline2} illustrates the distribution shown in Eq. \eqref{pnum1}, where $ n_{i} $ consists of two parts. The following results provides the expectation on $C_{0,0}$ and its variance, given fraction $\alpha$ of the $0-1$ altenate sequence.

\begin{theorem} \label{theorem.meanvarforCase1}
	The expectation of $ \operatorname{C}_{0,0}  $ is given by
	\begin{equation}
		\begin{aligned}
			\mathrm{E}\left(\operatorname{C}_{0,0}\right)
			=L \frac{\lambda_{s}}{2}[1-(1+\alpha) \varepsilon]+\varepsilon \frac{\lambda_{s}}{2}.\label{mean for case1.equation}
		\end{aligned}
	\end{equation}
     The expectation of variance of $ \operatorname{C}_{0,0}  $ is given by 
     \begin{align}
     	\mathrm{E}\left(\operatorname{Var}\left(\operatorname{C}_{0,0}\right)\right)=L\left(\frac{\lambda_{s}}{2}+\lambda_{b}\right).
     \end{align}
     \begin{proof}
     			See Appendix \ref{appendix.meanvarforCase1}.
   	 \end{proof}
\end{theorem}

\subsection{Case 2 : Estimate $\widehat{t_{start}}$ is chip $t_{0,k}$}
Assuming $t_{start} = \epsilon$, estimate $ t_{0,k} $ implies that correlation value $ C_{0,k} $ is the largest, where $ 1\leq k \leq n-1 $. Note that the calculation of $ C_{0,k} $ is the same for $ -(n-1)\leq k \leq -1 $, such that $k$ can be assumed as positive numbers. We have that

\begin{equation} \label{equ.correlation for case 2}
	\operatorname{C}_{0,k}  = \sum_{i=1}^{L}\left\{n_{i} \times\left(2 s_{i}-1\right)\right\}
\end{equation}	
where
\begin{equation}\label{pnum2}
	n_{i} \sim \left[\operatorname{Poi}\left((1-\frac{k}{n})\left(\lambda_{s} s_{i}+\lambda_{b}\right)\right)+\operatorname{Poi}\left(\varepsilon\left(\lambda_{s} s_{i}+\lambda_{b}\right)\right)+\operatorname{Poi}\left((\frac{k}{n}-\varepsilon)\left(\lambda_{s} s_{i+1}+\lambda_{b}\right)\right)\right].
\end{equation} 
\begin{figure}[htpb]
	\centering
	\includegraphics[width=6.5in]{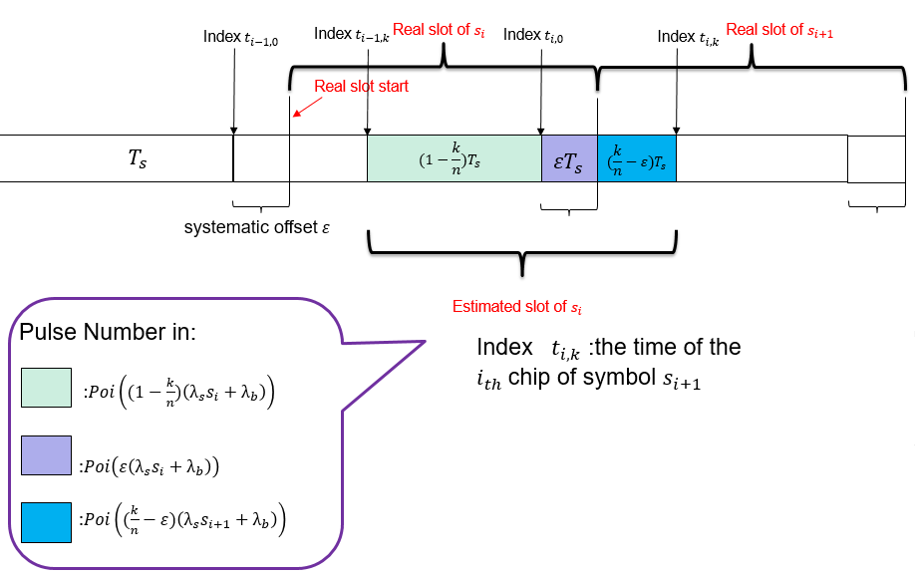}
	\caption{The timeline of the synchronization.} \label{timeline3}
\end{figure}
Fig. \ref{timeline3} illustrates the distribution shown in Eq. \eqref{pnum2}, where $ n_{i} $ consists of three parts.
\begin{theorem} \label{theorem.meanvarforCase2}
	The expectation of $ \operatorname{C}_{0,k}  $ is given by
	\begin{align} 
		\mathrm{E}\left(\operatorname{C}_{0,k}\right)=L \frac{\lambda_{s}}{2}[1-(1+\alpha) (\frac{k}{n}-\varepsilon)]+(\frac{k}{n}-\varepsilon)\frac{\lambda_{s}}{2}.\label{mean for case2.equation}
	\end{align}
   The expectation of variance of $ \operatorname{C}_{0,k}  $ is given by
\begin{align}
	\mathrm{E}\left(\operatorname{Var}\left(\operatorname{C}_{0,k}\right)\right)=L\left(\frac{\lambda_{s}}{2}+\lambda_{b}\right).
\end{align}
	\begin{proof}
		See Appendix \ref{appendix.meanvarforCase2}.
	\end{proof}
\end{theorem}

Consider the covariance between $ \operatorname{C}_{0,0}  $ and $ \operatorname{C}_{0,k}  $. Note that $ \operatorname{C}_{0,0}  $ re-expressed as follows,
\begin{equation} \label{pre_pnum3}
	\operatorname{C}_{0,0}  = \sum_{j=1}^{L}\left\{n_{j} \times\left(2 s_{j}-1\right)\right\},
\end{equation}
where
\begin{equation}\label{pnum3}
	n_{j} \sim \left[\operatorname{Poi}\left(\varepsilon\left(\lambda_{s} s_{j-1}+\lambda_{b}\right)\right)+\operatorname{Poi}\left((\frac{k}{n}-\varepsilon)\left(\lambda_{s} s_{j}+\lambda_{b}\right)\right)+\operatorname{Poi}\left((1-\frac{k}{n})\left(\lambda_{s} s_{j}+\lambda_{b}\right)\right)\right].
\end{equation} 
\begin{figure}[htpb]
	\centering
	\includegraphics[width=6.5in]{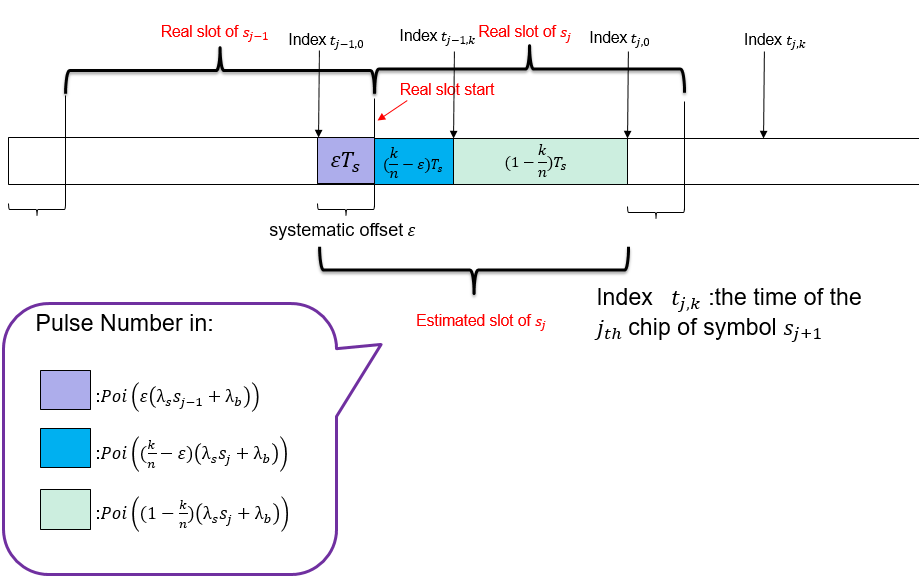}
	\caption{The timeline of the synchronization.} \label{timeline4}
\end{figure}
Fig. \ref{timeline4} illustrates the distributio shown in Eq. \eqref{pnum3}, where $ n_{j} $ consists of three parts.

The covariance is given by
\begin{equation}
	\operatorname{Cov}\left(\operatorname{C}_{0,0}, \operatorname{C}_{0, k}\right)=E\left\{\sum_{i=1}^{L} \sum_{j=1}^{L} n_{i} n_{j}\left(2 s_{i}-1\right)\left(2 s_{j}-1\right)\right\}-E\left(\operatorname{C}_{0,0}\right) E\left(\operatorname{C}_{0, k}\right).
\end{equation}

\begin{theorem}\label{theorem.covforCase2}
	 The expectation of covariance $ \operatorname{Cov}\left(\operatorname{C}_{0,0}, \operatorname{C}_{0, k}\right) $ is given by
	 \begin{align}\label{change0}
	 	\mathrm{E}\left\{\operatorname{Cov}\left(\operatorname{C}_{0,0}, \operatorname{C}_{0, k}\right)\right\}=L\left(\frac{\lambda_{s}}{2}+\lambda_{b}\right)\left[1-(1+\alpha) \frac{k}{n}\right]+\frac{k}{n}\left(\frac{\lambda_{s}}{2}+\lambda_{b}\right)-\varepsilon \frac{\lambda_{s}}{2}.
	 \end{align}
    	\begin{proof}
    	See Appendix \ref{appendix.covforCase2}.
    \end{proof}
\end{theorem}

\subsection{Case 3 : the Estimation $\widehat{t_{start}}$ is the Time of Chip with Index(2m,k)}
Similar to Case 1, assuming $t_{start} = \epsilon$, consider the estimate of timeIndex $ t_{2m,k} $, where correlation value $ C_{2m,k} $ is the largest, for $ m\geq 1  $ and $ -n\leq k \leq n-1 $. The calculation of $ C_{2m,k} $ is the same for $ m \leq -1 $, such that $m$ is assumed as positive for convenience.
\begin{equation} \label{equ.correlation for case 3}
	\operatorname{C}_{2m,k}  \sim \sum_{i=1}^{L}\left\{n_{i} \times\left(2 s_{i}-1\right)\right\},~~2m \leq \min\{\alpha L, \frac{1-\alpha}{2}L\}
\end{equation}	
where
\begin{equation}\label{pnum4}
	n_{i} \sim \left[\operatorname{Poi}\left((1-\frac{k}{n})\left(\lambda_{s} s_{i+2m}+\lambda_{b}\right)\right)+\operatorname{Poi}\left(\varepsilon\left(\lambda_{s} s_{i+2m}+\lambda_{b}\right)\right)+\operatorname{Poi}\left((\frac{k}{n}-\varepsilon)\left(\lambda_{s} s_{i+2m+1}+\lambda_{b}\right)\right)\right].
\end{equation} 
\begin{figure}[htpb]
	\centering
	\includegraphics[width=6.5in]{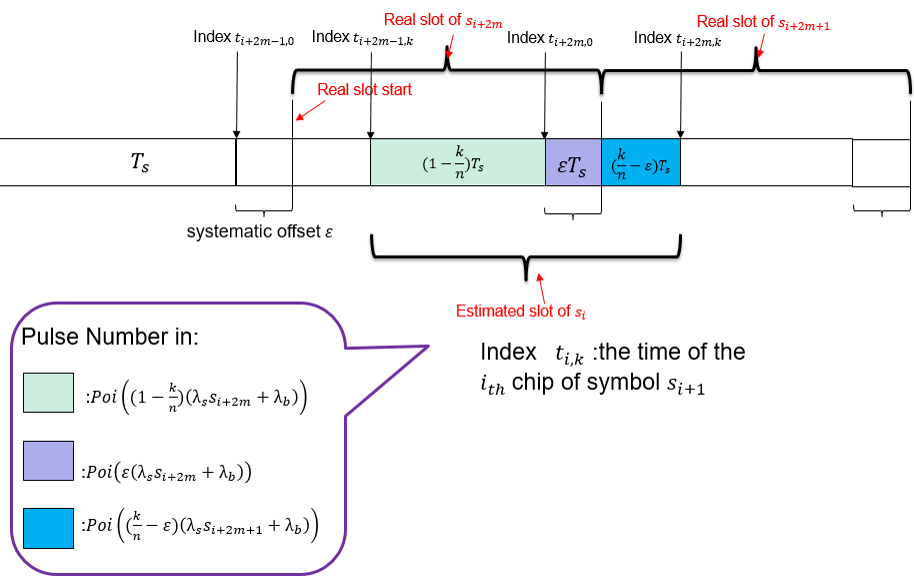}
	\caption{The timeline of the synchronization.} \label{timeline5}
\end{figure}
Fig. \ref{timeline5} illustrates the distribution as shown in Eq. \eqref{pnum4}, where $ n_{i} $ consists of three parts.

Different from Case 1 and Case 2, in Case 3, $ 2m $ symbols are overlapped between RANDOM part and $\{1,0,...,1,0\} $ part, as well as missed $ 2m $ symbols on the edge of entire synchronization sequence, cannot be ignored. We have the following results. 

\begin{theorem} \label{theorem.meanvarforCase3}
The expectation of  $ \operatorname{C}_{2m,k} $ is given by
\begin{align}
	\mathrm{E}\left(\operatorname{C}_{2m, k}\right)=\frac{\lambda_{s}}{2}\left\{(\alpha L-2 m)\left[1-2\left(\frac{k}{n}-\varepsilon\right)\right]+\left(\frac{k}{n}-\varepsilon\right)\right\},~~2m \leq \min\{\alpha L, \frac{1-\alpha}{2}L\}.
	\label{mean for case3.equation}
\end{align}

The expectation of variance of $ \operatorname{C}_{2m,k}  $ is given by
\begin{align}
	\mathrm{E}\left(\operatorname{Var}\left(\operatorname{C}_{2m,k}\right)\right)=\left(L-2m\right)\left(\frac{\lambda_{s}}{2}+\lambda_{b}\right),~~2m \leq \min\{\alpha L, \frac{1-\alpha}{2}L\}.
\end{align}

\begin{proof}
	See Appendix \ref{appendix.meanvarforCase3}.
\end{proof}
\end{theorem}

Considering the expression of $ \operatorname{C}_{0,0} $ in Eq.\eqref{pre_pnum3}, \eqref{pnum3}, the coveriance can be expressed as 
\begin{equation}
	\operatorname{Cov}\left(\operatorname{C}_{0,0}, \operatorname{C}_{2m, k}\right)=E\left\{\sum_{i=1}^{L} \sum_{j=1}^{L} n_{i} n_{j}\left(2 s_{i}-1\right)\left(2 s_{j}-1\right)\right\}-E\left(\operatorname{C}_{0,0}\right) E\left(\operatorname{C}_{2m, k}\right).
\end{equation}

\begin{theorem}\label{theorem.covforCase3}
The expectation of covariance is given by

\begin{align}
	&\mathrm{E}\left\{\operatorname{Cov}\left(\operatorname{C}_{0,0}, \operatorname{C}_{2m, k}\right)\right\}=\left(\frac{\lambda_{s}}{2}+\lambda_{b}\right)\left[\left(\alpha L -2m \right)\left(1-2\frac{k}{n}\right)\right]+\frac{k}{n}\left(\frac{\lambda_{s}}{2}+\lambda_{b}\right)-\varepsilon\frac{\lambda_{s}}{2},
\end{align}
where $2m \leq \min\{\alpha L, \frac{1-\alpha}{2}L\} $.
\begin{proof}
	See Appendix \ref{appendix.covforCase3}.
\end{proof}
\end{theorem}

\section{Estimation of the Synchronization Accuracy} \label{section.estimation of the synchronization accuracy}
When estimating starting time $ \widehat{t_{start}} $ as $ t_{m,k} $, we have that $ \operatorname{C}_{m, k} \geq \operatorname{C}_{tIndex}, \forall tIndex \neq (m,k)  $.

The probability of such event, denoted as $p_{m,k}$, satisfying the following upper bound
\begin{align}
	p_{m, k}=P\left(\widehat{t_{\text {start }}}=t_{m, k}\right)=P\left(\operatorname{C}_{m, k} \geq \max_{\forall tIndex \neq (m,k)}~ \operatorname{C}_{\text {tIndex }}\right) \leq P\left(\operatorname{C}_{m, k} \geq \operatorname{C}_{0,0}\right) .
\end{align}

The above probability can be approximated based on $ \mathrm{E}\left(\operatorname{C}_{m, k}\right)) $, $\operatorname { V a r }\left(\operatorname{C}_{m, k}\right) $ and $ \operatorname{Cov}\left(\operatorname{C}_{0,0}, \operatorname{C}_{0, k}\right) $.

\subsection{Gaussian Approximation on the the Correlation Value}
Consider a simple case zero systematic error $ \varepsilon $, where the correlation value can be expressed as the difference between sum of two sets of independently identically distributed random variables,
\begin{equation}
	\begin{aligned}
		\operatorname{C}=\sum_{i \in \left\{i : s_{i} =1\right\}}x_{i}-\sum_{i \in \left\{i : s_{i} =0\right\}}y_{i},\\
		x_{i} \stackrel{i.i.d}{\sim} Poi\left(\lambda_{s}+\lambda_{b}\right),~ y_{i} \stackrel{i.i.d}{\sim}Poi\left(\lambda_{b}\right).
	\end{aligned}
\end{equation}
Note that for length $ L \geq 64$, $ \sum_{i \in \left\{i : s_{i} =1\right\}}x_{i} $ and $ \sum_{i \in \left\{i : s_{i} =0\right\}}y_{i} $ can be approximated by Gaussian distribution, and linear combination of Gaussian random variables also satisfies Gaussian distribution. As a consequence, we assume that $ \operatorname{C}_{m, k} $ can be approximated by Gaussian distribution $ {\cal N}(\mathrm{E}\left(\operatorname{C}_{m, k}\right)),\operatorname { V a r }\left(\operatorname{C}_{m, k}\right)) $.

\begin{align}
	P\left(\operatorname{C}_{m, k} \geq \operatorname{C}_{0,0}\right) = P\left(\operatorname{C}_{0, 0}-\operatorname{C}_{m,k} \leq 0\right).
\end{align}
The distribution of $ \left(\operatorname{C}_{0, 0}-\operatorname{C}_{m,k}\right) $, as the linear combination of Gaussian random variables, satisfies the following Gaussian distribution,
\begin{equation}\label{gaussian}
	\begin{aligned}
		&\left( \operatorname{C}_{0, 0}-\operatorname{C}_{m,k}\right) \\
		&\sim {\cal N}(\left(\mathrm{E}\left(\operatorname{C}_{0, 0}\right)-\mathrm{E}\left(\operatorname{C}_{m,k}\right)\right),\left\{\operatorname { V a r }\left(\operatorname{C}_{0, 0}\right)+\operatorname { V a r }\left(\operatorname{C}_{m, k}\right)-2\times \operatorname{Cov}\left(\operatorname{C}_{0,0}, \operatorname{C}_{m, k}\right)\right\}).
	\end{aligned}
\end{equation}
%

\subsection{Upper Bound on the Probability of the Offset}

Based on the Gaussian approximation, an upper bound on $p_{m,k}$ can be expressed as follows,
\begin{align}\label{pmk}
	p_{m,k} \leq \bar{p_{m, k}}=\Phi\left(\frac{0-\left(\mathrm{E}\left(\operatorname{C}_{0, 0}\right)-\mathrm{E}\left(\operatorname{C}_{m, k}\right)\right)}{\sqrt{\operatorname { V a r }\left(\operatorname{C}_{0, 0}\right)+\operatorname { V a r }\left(\operatorname{C}_{m, k}\right)-2\times \operatorname{Cov}\left(\operatorname{C}_{0,0}, \operatorname{C}_{m, k}\right)}}\right).
\end{align}
In particular, we have
\begin{equation}
p_{1, 0} \leq P\left(\operatorname{C}_{0, 0} \geq \operatorname{C}_{0,1}\right) = 1-P\left(\operatorname{C}_{0, 1} \geq \operatorname{C}_{0,0}\right).
\end{equation} 
For Case 1, we have
\begin{equation}
	\bar{p_{0, 0}}=1-\bar{p_{0, 1}}.
\end{equation}

\begin{theorem}\label{theorem.UBforPro}
    For Case 2, an upper bound on probability of $ \hat{t_{start}} = t_{0,k} $ is given by
    \begin{align}
    	p_{0, k} \leq \bar{p_{0, k}}=\Phi\left(\frac{\frac{\lambda_{s}}{2}\left(2\varepsilon-\frac{k}{n}\right)\left[L\left(1+\alpha\right)-1\right]}{\sqrt{2\frac{k}{n}\left(\frac{\lambda_{s}}{2}+\lambda_{b}\right)\left[\left(1+\alpha\right)L-1\right]+\varepsilon\lambda_{s}}}\right).
    \end{align}
    For Case 3, an upper bound on probability of $ \hat{t_{start}} = t_{2m,k} $ is given by
    \begin{align}
    	p_{2m, k} \leq \bar{p_{2m,k}}=\Phi\left(\frac{\left(\alpha L-2m\right)\frac{\lambda_{s}}{2}\left[1-2\left(\frac{k}{n}-\varepsilon\right)\right]-L\frac{\lambda_{s}}{2}\left[1-\left(1+\alpha\right)\varepsilon\right]-\left(2\varepsilon-\frac{k}{n}\right)\frac{\lambda_{s}}{2}}{\sqrt{2\left(\frac{\lambda_{s}}{2}+\lambda_{b}\right)\left[\left(L-m\right)-\left(\alpha L-2m\right)\left(1-2\frac{k}{n}\right)-\frac{k}{n}\right]+\varepsilon\lambda_{s}}}\right).
    \end{align}
    \begin{proof}
    	The above results can be proved by taking the expectation results on $ \mathrm{E}\left(\operatorname{C}_{m, k}\right)) $, $\operatorname { V a r }\left(\operatorname{C}_{m, k}\right) $ and $ \operatorname{Cov}\left(\operatorname{C}_{0,0}, \operatorname{C}_{0, k}\right) $. The detailed process is standard and thus not given here.
    \end{proof}
\end{theorem}

\subsection{Upper Bound on the Offset}

We calculate the upper bound on the mean squared offset based on the upper bounds on the probability. Given initial offset $\epsilon$, the error is denoted as $ e=\widehat{t_{start}}-\epsilon $. By analysis above, $\hat{t_{start}}$ can be $t_{m,k}$, and $e_{m,k}= t_{m,k} - \epsilon $. The mean squared error, denoted as  $\mathrm{E}\left(e^{2} \mid \varepsilon\right) $ can be expressed as,
	\begin{equation}
		\begin{aligned}
			\mathrm{E}\left(e^{2} \mid \varepsilon\right)&=\sum_{m, k} e_{m, k}^{2} \times p_{m, k}\\
			&=e_{0,0}^{2} \times p_{0,0}+2 \sum_{k=1}^{n-1} e_{0, k}^{2} \times p_{0, k}+2 \sum_{m=1}^{\infty} \sum_{k=-(n-1)}^{n} e_{2 m, k}^{2} \times p_{2 m, k},
		\end{aligned}
	\end{equation}
where $ \varepsilon $ meets uniform distribution $ U(-\frac{T_c}{2} ,\frac{T_c}{2}) $ as a result of systematic offset. We have the following result.

\begin{theorem}
	Based on Gaussian approximation, an upper bound on $ e^2 $ is given by
	\begin{align}
		\mathrm{E}\left(\bar{e^{2}}\right)=\int_{-\frac{T_c}{2}}^{\frac{T_c}{2}}\left(	e_{0,0}^{2} \times \bar{p_{0,0}}+2 \sum_{k=1}^{n-1} e_{0, k}^{2} \times \bar{p_{0, k}}+2 \sum_{m=1}^{\infty} \sum_{k=-(n-1)}^{n} e_{2 m, k}^{2} \times \bar{p_{2 m, k}}\right) p(\varepsilon) d \varepsilon,
	\end{align}
     where $ p\left(\varepsilon\right) = \frac{1}{\frac{T_c}{2}-\left[-\frac{T_c}{2}\right]} = \frac{1}{T_c},~~ \varepsilon \in \left[-\frac{T_c}{2},\frac{T_c}{2}\right] $.
\end{theorem}

\section{Optimization of the Synchronization Sequence} \label{section.optimization of the synchronization sequence}

\subsection{Optimization Objective}
The objective is to minimize MSE $ \mathrm{E}\left(\bar{e^{2}}\right) $ subject to the constraint the probability of symbol-level offset is below a low threshold.

Note that as $ \alpha $ increases, the probability of $ \widehat{t_{start}}=t_{0,0} $ increases, the probability of $ \widehat{t_{start}}=t_{0,k} $ decreases and the probability of $ \widehat{t_{start}}=t_{2m,k} $ increases. The probability of $ \widehat{t_{start}}=t_{2m,k} $ needs to be sufficiently small to avoid significant performance loss of signal detection, which translates to constraint that $ 1\leq \alpha \textless\alpha_{threshold} $. In general, $\mathrm{E}\left(\bar{e^{2}} \right)$ mostly decreases first and increases later with $ \alpha $. 

In particular, we characterize a more rigorous setting of $ \alpha $ to avoid Case 3. The upper bound on the probability of Case 3 need be lower than threshold $\eta$, which should be sufficiently small, such as $ 10^{-8} $, which can be expressed as
\begin{align}\label{premise}
	P\left(Case~3\right) =2 \sum_{m=1}^{\infty} \sum_{k=-(n-1)}^{n} \bar{p_{2 m, k}} \leq \eta.
\end{align}

\subsection{Mathematical Expression of the Optimization}
\begin{theorem}\label{opm}
	Optimization of the synchronization sequence is committed to minimize MSE $\mathrm{E}\left(\bar{e^{2}} \right)$ under the constraint condition that the upper bound on the probability of Case 3 need be lower than threshold $\eta$, which can be expressed as
   	\begin{equation}
   	\begin{aligned} \label{P}
   		&minimize \quad \int_{-\frac{T_c}{2}}^{\frac{T_c}{2}}\left(	e_{0,0}^{2} \times \bar{p_{0,0}}+2 \sum_{k=1}^{n-1} e_{0, k}^{2} \times \bar{p_{0, k}}+2 \sum_{m=1}^{\infty} \sum_{k=-(n-1)}^{n} e_{2 m, k}^{2} \times \bar{p_{2 m, k}}\right) p(\varepsilon) d \varepsilon\\
   		&\begin{array}{r@{\quad}r@{}l@{\quad}l}
   			subject ~to &	2 \sum_{m=1}^{\infty} \sum_{k=-(n-1)}^{n} \bar{p_{2 m, k}} -\eta \leq 0, \\
   			            &   -\alpha \leq 0,\\
   			            &   \alpha-1 \leq 0.
   		\end{array}
   	\end{aligned}
   \end{equation}
\end{theorem}
First, we need to find the interval $ [0,\alpha_{threshold}] $ subject to $ P\left(Case~3 \textbar \alpha \in [0,\alpha_{threshold}] \right)\leq \eta $.
\begin{algorithm}[H]
	\caption{Search Method for the interval $ [0,\alpha_{threshold}] $}
	\label{alg1}
	\begin{algorithmic}[1]
		\Require $ P(Case~3 \textbar \alpha)$, the search accuracy $ h $
		\Ensure $\alpha_{threshold}$ \\ 
		
		\textbf{Initialize:$ a=0,b=1 $}
		
		\While{$(b \geq a)$ and $( g \textgreater \eta) $ }\\
		$ \alpha_{1}=b-h $,$ g =  P(Case~3 \textbar \alpha=\alpha_{1})$
		\EndWhile 
		\State $\alpha_{threshold}= \alpha_{1}$.
		
	\end{algorithmic}
\end{algorithm}

We perform Golden Search Method on the interval $ [0,\alpha_{threshold}] $ to find the minimum. 
\begin{algorithm}[H]
	\caption{Golden Search Method for the optimization problem \ref{P}}
	\label{alg2}
	\begin{algorithmic}[1]
		\Require $\alpha_{threshold}$, $\mathrm{E}\left(\bar{e^{2}} \right)\left(\alpha\right)$, the search accuracy $ h $
		\Ensure $\alpha_{0}$ \\ 
		
		\textbf{Initialize:$ a=0,b=\alpha_{threshold} $}\\
		The insertion point generated by using the golden section method: \\
		 $ \alpha_{1} = a + 0.382(b-a)$,
	   	$ \alpha_{2} = a + 0.618(b-a) $
		\While{$\left(b-a \geq h\right)$}
		\State $f_{1}= \mathrm{E}\left(\bar{e^{2}} \right)\left(\alpha_{1}\right)$,$f_{2}= \mathrm{E}\left(\bar{e^{2}} \right)\left(\alpha_{2}\right)$
		   \If{$f_{1} \textgreater f_{2}$}
		          \State  $ a=\alpha_{1},\alpha_{1}=\alpha_{2},\alpha_{2}=a+0.618(b-a) $
		         
		          \Else
		          \State $b=\alpha_{2},\alpha_{2}=\alpha_{1},\alpha_{1}=a+0.382(b-a) $
		          \EndIf
		\EndWhile 
		\State $\alpha_{0}= \frac{\alpha_{1}+\alpha_{2}}{2}$.
		
	\end{algorithmic}
\end{algorithm}

\section{Verification by Simulating} \label{section.verification by simulating}

There are simulation experiments to verify the derivation in Section \ref{section.Mathematical expectation,variance and covariance of correlation values}, Section \ref{section.estimation of the synchronization accuracy}  and Section \ref{section.optimization of the synchronization sequence}.

The simulation for Subsection \ref{veriStat} and \ref{veriG} includes two groups. The number of chips in one symbol satisfies $ n=100 $ and other parameters' setting is,
\begin{center}
	\captionof{table}{The parameters' setting for simulating}
	\label{table.para1}
	\begin{tabular}{| c | c | c | c | c |}
		\hline
		Group &Sequence length $ L $ & $ alpha $ & Signal intensity $ \lambda_{s}$ & Noise intensity $ \lambda_{b}$ \\ \hline
		$ 1 $ & $ 128 $ &$  0.707 $ &$ 10 $&$ 1 $\\ \hline
		$ 2 $ & $ 256 $ &$  0.853 $ &$ 8 $&$ 1 $\\
		\hline
	\end{tabular}
\end{center}

We simulate the Poisson process for arrival of photons. Treated $ t_{m,k} $ as the starting time, we divide the time slots for each symbol $ s_{i} $. $ C_{m,k} $ can be figured up by correlation with synchronous sequence. The above steps are repeated  repeated $10,000$ times. Then figure out the mean and variance of $ C_{m,k} $. And we can get the covariance of $ C_{m,k} $ and $ C_{0,0} $

\subsection{Verification for Statistical Analysis of Correlation Value}\label{veriStat}
After $ 10,000 $ times repetition, we need figure out the mean and variance of $ C_{m,k} $. And we can get the covariance of $ C_{m,k} $ and $ C_{0,0} $. The timeline is from $ -10.00 $ to $ 10.00 $, on which $ t_{m,k} $ are discrete points with $ t_{-10,0}=-10.00 $, $ t_{0,-1}=-0.01 $, $ t_{0,0}=0.00 $, $ t_{0,1}=0.01 $, $ t_{0,2}=0.02 $, $t_{1,1}=1.01$ and so on. ( $T_s=1$  is noted as a unit time.)

\begin{figure}[htbp]
	\centering
	\includegraphics[width=5.0in]{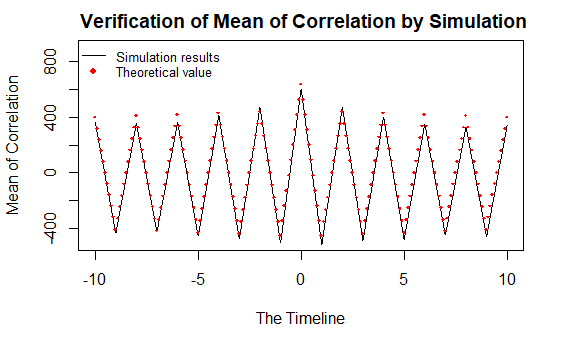}
	\caption{Comparison of theoretical and simulation results on correlation value's mean for Group 1} \label{128mean}
\end{figure}

\begin{figure}[htbp]
	\centering
	\includegraphics[width=5.0in]{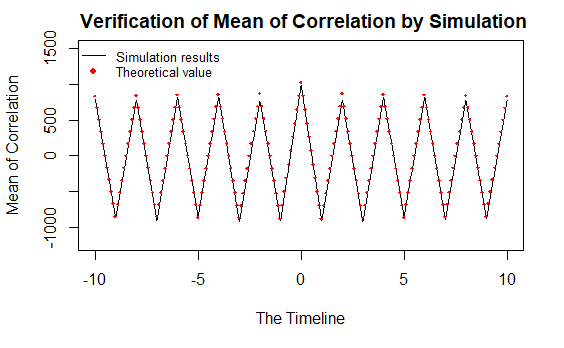}
	\caption{Comparison of theoretical and simulation results on correlation value's mean for Group 2} \label{256mean}
\end{figure}


The mean of correlation value is shown in Fig \ref{128mean} and Fig \ref{256mean}. The black line charts represent the average of $ 10,000 $ correlation values when the time on the horizontal axis is considered as the starting time. As we know, the equations \ref{mean for case1.equation}, \ref{mean for case2.equation} and \ref{mean for case3.equation} show correlation values' mathematical expectations of a group of different synchronous sequences with the same length and the same parts' ratio $ \alpha $. The sysmetric offset $ \varepsilon $ is a random variable that meets uniform distribution here. So we need do one more step which is,
\begin{align}
	\mathrm{E}\left\{\mathrm{E}\left(C_{m,k} \mid \varepsilon\right)\right\}=\int_{-\frac{T_c}{2}}^{\frac{T_c}{2}}\mathrm{E}\left(C_{m,k} \mid \varepsilon\right)p(\varepsilon) d \varepsilon,
\end{align}
Then we get the theoretical values represented by the red points in Fig \ref{128mean} and Fig \ref{256mean}. For $ 10,000 $ times repetition, the synchronous sequence is fixed. So the theoretical values actually are mathematical expectations of simulation results. The simulation results are not only close to their theoretical expectations in value, but also show consistent increase or decrease trend.

\begin{figure}[htbp]
	\centering
	\includegraphics[width=5.0in]{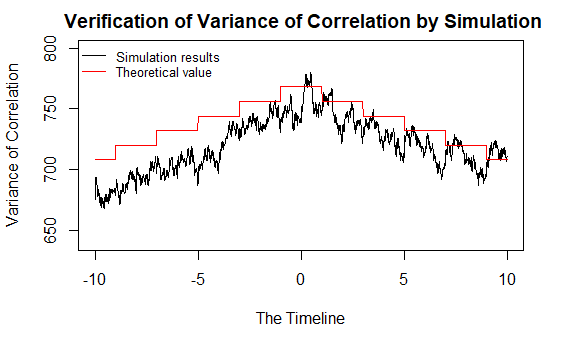}
	\caption{Comparison of theoretical and simulation results on correlation value's variance for Group 1} \label{128var}
\end{figure}

\begin{figure}[htbp]
	\centering
	\includegraphics[width=5.0in]{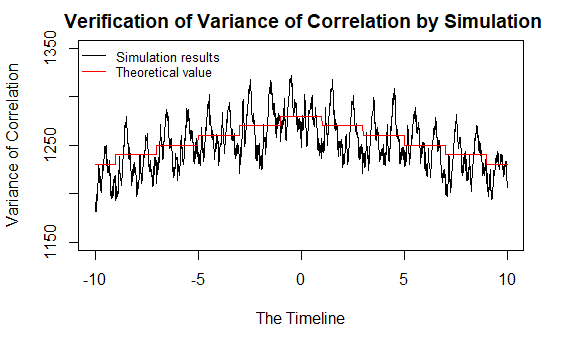}
	\caption{Comparison of theoretical and simulation results on correlation value's variance for Group 2} \label{256var}
\end{figure}

The variance of correlation value is shown in Fig \ref{128var} and Fig \ref{256var}.The black line charts represent the sample variance of $ 10,000 $ correlation values while the red line charts represent their theoretical mathematical expectations. The simulation results fluctuate around the theoretical expectations. In particular, Group 2 shows better consistency due to its sequence structure, such longer $ L $ and larger $ \alpha $. The theoretical value is approximately the fluctuation center line of the simulation results. This makes the theoretical value relatively instructive in Group 2.

\begin{figure}[htbp]
	\centering
	\includegraphics[width=5.0in]{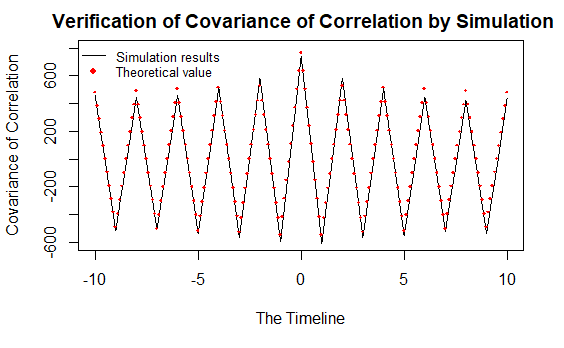}
	\caption{Comparison of theoretical and simulation results on correlation value's covariance for Group 1} \label{128cov}
\end{figure}

\begin{figure}[htbp]
	\centering
	\includegraphics[width=5.0in]{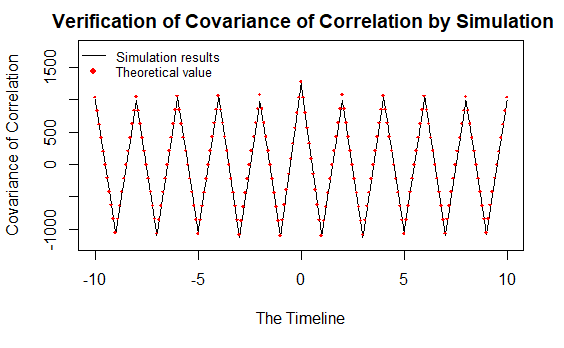}
	\caption{Comparison of theoretical and simulation results on correlation value's covariance for Group 2} \label{256cov}
\end{figure}

The covariance of correlation value is shown in Fig \ref{128cov} and Fig \ref{256cov}. The black line charts represent the sample covariance of correlation values while the red points represent their theoretical mathematical expectations. The great consistency both in value and changing trends is shown.

\subsection{Verification for Gaussian Assumption }\label{veriG}
Kolmogorov-Smirnov test is used to explore if the samples satisfy Gaussian distribution. For each Group, $ 10,000 $ samples of $ C_{m,k} $ is the test set and p-value $ p-value_{m,k} $ displays the result. When $ p-value_{m,k} \leq 0.05 $, reject the Gaussian hypothesis. Otherwise, accept the Gaussian hypothesis. There are $ 2,001 $ time points $ t_{m,k} $. The result is shown in table \ref{table.ks}.
\begin{center}
	\captionof{table}{The results of KS test}
	\label{table.ks}
	\begin{tabular}{| c | c | c | c | c |}
		\hline
		Group &Accepted & Rejected &In total & mean of p-value\\ \hline
		$ 1 $ & $ 1476 $ &$  525 $ &$ 2001 $& $ 0.1039499 $\\ 
		\hline
		$ 2 $ & $ 1781 $ &$  220 $ &$ 2001 $& $ 0.1940755 $\\ 
		\hline
	\end{tabular}
\end{center}
Most hypotheses are accepted. Means of p-value are more than $ 0.05 $. It is shown that Gaussian assumption is reliable.
\subsection{Verification for the Selection of $ \alpha $ }\label{veriA}
Combining optimization Theorem \ref{opm}, $ \lambda_{s} $ and $ \lambda_{b} $ are taken from the worst signal-to-noise condition of the current scenario to avoid failed synchronization. The number of chips in one symbol is decided by the clock sampling frequency of the hardware. And extremely small $ p_{2m,k}$ can be ignored when $m \textgreater 10 $. Set $ \lambda_s = 5 $, $ \lambda_b = 1 $ as the worst signal-to-noise condition, and the number of chips in one symbol satisfies $ n=100 $.  
We have the optimization result as table \ref{table.alpha}
\begin{center}
	\captionof{table}{The optimization of $ \alpha $ for synchronization sequences}
	\label{table.alpha}
	\begin{tabular}{| c | c | c | c |}
		\hline
		Sequence length $ L $ & $ \alpha_{0}=\mathop{\arg\min}_{\alpha}\mathrm{E}\left(\bar{e^{2}} \right) $ &
		$ \left\{\alpha \colon 2 \sum_{m=1}^{\infty} \sum_{k=-n}^{n} \bar{p_{2 m, k}} \leq 10^{-8}\right\} $ &
		final $ alpha_{chosen} $\\ \hline
		$ 128 $ & $ 0.770 $ &$  \left\{\alpha \colon \alpha \leq 0.719\right\} $ &$ 0.719 $\\ \hline
		$ 256 $ & $ 0.863 $ &$  \left\{\alpha \colon \alpha \leq 0.853\right\} $ &$ 0.853 $\\ \hline
		$ 512 $ & $ 0.917 $ &$  \left\{\alpha \colon \alpha \leq 0.913\right\} $ &$ 0.913 $\\ 
		\hline
	\end{tabular}
\end{center}

Then we can get the synchronization sequence which is both reliable and has a high synchronization accuracy with final $ alpha_{chosen} $ corresponding to sequence length. 
\begin{figure}[htbp]
	\centering
	\includegraphics[width=5.0in]{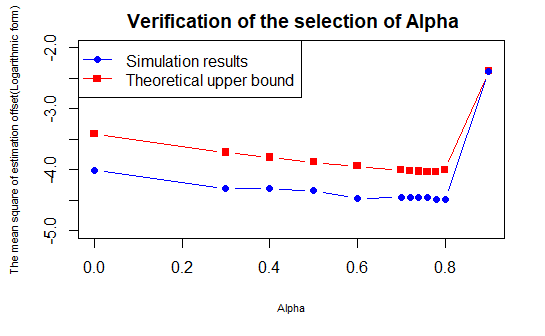}
	\caption{Comparison of theoretical and simulation results on the mean square of estimation offset for Group 1} \label{128alpha}
\end{figure}

\begin{figure}[htbp]
	\centering
	\includegraphics[width=5.0in]{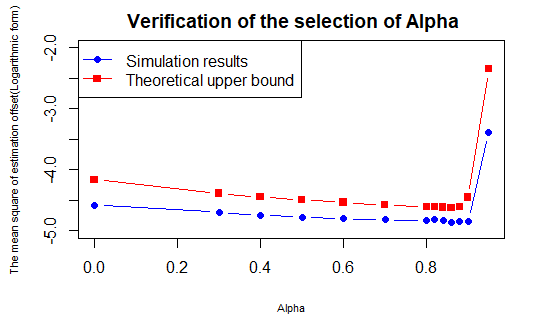}
	\caption{Comparison of theoretical and simulation results on the mean square of estimation offset for Group 2} \label{256alpha}
\end{figure}
There are also two groups' simulation to verificate the optimization result. For Group 1, all the synchronization sequences' length is $ L=128 $. The proportion of $\{1,0,...\} $ part ($ \alpha $) is different. The result is shown in Fig \ref{128alpha}. In particular, the vertical axis is logarithmic base 10. Similarly, 
Fig \ref{256alpha} is for synchronization sequences' length $ L=256 $ in Group 2. The average of $ e^2 $ in simulation results is less than corresponding theoretical upper bounds. Mostly, their orders of magnitude are the same. As for the lowest points in simulation results and the corresponding theoretical upper bounds, their horizontal axis coordinates are very close. All these demonstrate the reliability of the derivation of the theoretical upper bound, and thus obtain a reasonable $ \alpha $ selection.



From the specific simulation results in Group 1, $ e^2=0.0000344911589103836 $ with $ \alpha=0.72 \approx 0.719 $ (refer to table \ref{table.alpha}), while $ e^2=0.0000974964953102525 $ when the sequence is completely randomly generated with $ \alpha=0 $. The synchronization accuracy has been increased by about three times. In Group 2, $ e^2=0.0000136843212259768 $ with $ \alpha=0.86 \approx 0.853 $ (refer to table \ref{table.alpha}), while $ e^2=0.0000263214154550494 $ when the sequence is completely randomly generated with $ \alpha=0 $. The synchronization accuracy has been increased by about two times. This shows that the design and optimization of synchronization sequence can obtain a significant advance in accuracy compared with random sequence.

\section{Conclusion} \label{section.conclusion}
In Poisson channel of the ultraviolet (UV) scattering communication, the synchronization is based on maximum Pulse Number-Sequence correlation problem. According to the calculated properties of correlation values, we design the synchronization sequence including two equilong RANDOM parts (Symbols meet Bernoulli distribution with equal probability.) and a $\{1,0,1,0,1,0,...,1,0,1,0\}$ part between them with $ \alpha $ as its proportion of entire sequence. Based on Gaussian hypothesis and derivation of Poisson's process, we quantify the relationship of synchronization's accuracy and $ \alpha $ to optimize the synchronization sequence. The simulation results show that the derivation is reasonable. And the optimization of synchronization sequence results in a outstanding improvement in accuracy, which is of great significance to channel estimation, improvement of codec accuracy.

\begin{appendices} 
	\section{ Proof of Theorem \ref{theorem.meanvarforCase1} }  \label{appendix.meanvarforCase1}
Note that
	\begin{equation} 
		\operatorname{C}_{0,0}  \sim \sum_{i=1}^{L}\left\{n_{i} \times\left(2 s_{i}-1\right)\right\},
	\end{equation}
where
	\begin{equation}
		n_{i} \sim \left[\operatorname{Poi}\left(\varepsilon\left(\lambda_{s} s_{i-1}+\lambda_{b}\right)\right)+\operatorname{Poi}\left((1-\varepsilon)\left(\lambda_{s} s_{i}+\lambda_{b}\right)\right)\right].
	\end{equation} 
Firstly, if $ s_i$ lies in RANDOM part, we have $ \mathrm{E}\left(s_{i-1}s_{i}\right) = \frac{1}{4}$ when its two neighbors also lie in RANDOM part, and thus
	\begin{equation}
		\begin{aligned}
			&\mathrm{E}\left\{\operatorname{Poi}\left(\varepsilon\left(\lambda_{s} s_{i-1}+\lambda_{b}\right)\right) \times\left(2 s_{i}-1\right)\right\}  \\
			&=\mathrm{E}\left\{\mathrm{E}\left[\left(\varepsilon\left(\lambda_{s} s_{i-1}+\lambda_{b}\right)\right) \times\left(2 s_{i}-1\right) \mid \bm{s}\right]\right\} \\
			&=\varepsilon\mathrm{E}\left\{2\lambda_{s} s_{i-1}s_{i}-\lambda_{s} s_{i-1}+2\lambda_{b}s_{i}-\lambda_{b}\right\}\\
			&=0,
		\end{aligned}
	\end{equation}

    \begin{equation}
    	\begin{aligned}
    		&\mathrm{E}\left\{\operatorname{Poi}\left(\left(1-\varepsilon\right)\left(\lambda_{s} s_{i}+\lambda_{b}\right)\right) \times\left(2 s_{i}-1\right)\right\} \\
    		&=\mathrm{E}\left\{\mathrm{E}\left[\left(\left(1-\varepsilon\right)\left(\lambda_{s} s_{i}+\lambda_{b}\right)\right) \times\left(2 s_{i}-1\right)\mid \bm{s}\right]\right\} \\
    		&=\left(1-\varepsilon\right)\mathrm{E}\left\{2\lambda_{s}s_{i}^2-\lambda_{s}s_{i}+2\lambda_{b}s_{i}-\lambda_{b}\right\}\\
    		&=(1-\varepsilon) \frac{\lambda_{s}}{2}.
    	\end{aligned}
    \end{equation}
If $ s_i$ lies in RANDOM part, we have $ s_{i-1}=0$ when the boundary is between  $ s_{i-1}$ and  $ s_{i}$ or $ i=1 $, and thus
\begin{equation}
	\begin{aligned}
		&\mathrm{E}\left\{\operatorname{Poi}\left(\varepsilon\left(\lambda_{s} s_{i-1}+\lambda_{b}\right)\right) \times\left(2 s_{i}-1\right)\right\}  \\
		&=\varepsilon\mathrm{E}\left\{2\lambda_{s} s_{i-1}s_{i}-\lambda_{s} s_{i-1}+2\lambda_{b}s_{i}-\lambda_{b}\right\}\\
		&=0,
	\end{aligned}
\end{equation}

\begin{equation}
	\begin{aligned}
		&\mathrm{E}\left\{\operatorname{Poi}\left(\left(1-\varepsilon\right)\left(\lambda_{s} s_{i}+\lambda_{b}\right)\right) \times\left(2 s_{i}-1\right)\right\} \\
		&=\left(1-\varepsilon\right)\mathrm{E}\left\{2\lambda_{s}s_{i}^2-\lambda_{s}s_{i}+2\lambda_{b}s_{i}-\lambda_{b}\right\}\\
		&=(1-\varepsilon) \frac{\lambda_{s}}{2}.
	\end{aligned}
\end{equation}

If $ s_i $ lies in $\{1,0,...\} $ part, we have$ \mathrm{E}\left(s_{i-1}s_{i}\right) = 0$ when its two neighbors also lie in $\{1,0,...\} $ part. For $ (\alpha L - 2) $ items which don't lie in boundaries, we have $ \mathrm{E}\left(s_{i}\right) =\frac{1}{2}$ and $ \mathrm{E}\left(s_{i-1}\right) =\frac{1}{2}$.
	\begin{equation}
		\begin{aligned}
			&\mathrm{E}\left\{\operatorname{Poi}\left(\varepsilon\left(\lambda_{s} s_{i-1}+\lambda_{b}\right)\right) \times\left(2 s_{i}-1\right)\right\}  \\
			&=\mathrm{E}\left\{\mathrm{E}\left[\left(\varepsilon\left(\lambda_{s} s_{i-1}+\lambda_{b}\right)\right) \times\left(2 s_{i}-1\right) \mid \bm{s}\right]\right\} \\
			&=\varepsilon\mathrm{E}\left\{2\lambda_{s} s_{i-1}s_{i}-\lambda_{s} s_{i-1}+2\lambda_{b}s_{i}-\lambda_{b}\right\}\\
			&=(-\varepsilon) \frac{\lambda_{s}}{2},
		\end{aligned}
	\end{equation}
	
	\begin{equation}
		\begin{aligned}
			&\mathrm{E}\left\{\operatorname{Poi}\left(\left(1-\varepsilon\right)\left(\lambda_{s} s_{i}+\lambda_{b}\right)\right) \times\left(2 s_{i}-1\right)\right\} \\
			&=\mathrm{E}\left\{\mathrm{E}\left[\left(\left(1-\varepsilon\right)\left(\lambda_{s} s_{i}+\lambda_{b}\right)\right) \times\left(2 s_{i}-1\right)\mid \bm{s}\right]\right\} \\
			&=\left(1-\varepsilon\right)\mathrm{E}\left\{2\lambda_{s}s_{i}^2-\lambda_{s}s_{i}+2\lambda_{b}s_{i}-\lambda_{b}\right\}\\
			&=(1-\varepsilon) \frac{\lambda_{s}}{2}.
		\end{aligned}
	\end{equation}
If $ s_i$ lies in $\{1,0,...\} $ part, we have $ s_{i-1}$ lies in RANDOM part and $ s_i=1 $ when the boundary is between  $ s_{i-1}$ and $ s_{i}$, and thus
	\begin{equation}
	\begin{aligned}
		&\mathrm{E}\left\{\operatorname{Poi}\left(\varepsilon\left(\lambda_{s} s_{i-1}+\lambda_{b}\right)\right) \times\left(2 s_{i}-1\right)\right\}  \\
		&=\varepsilon\mathrm{E}\left\{2\lambda_{s} s_{i-1}s_{i}-\lambda_{s} s_{i-1}+2\lambda_{b}s_{i}-\lambda_{b}\right\}\\
		&=\varepsilon( \frac{\lambda_{s}}{2}+\lambda_{b}),
	\end{aligned}
\end{equation}

\begin{equation}
	\begin{aligned}
		&\mathrm{E}\left\{\operatorname{Poi}\left(\left(1-\varepsilon\right)\left(\lambda_{s} s_{i}+\lambda_{b}\right)\right) \times\left(2 s_{i}-1\right)\right\} \\
		&=\left(1-\varepsilon\right)\mathrm{E}\left\{2\lambda_{s}s_{i}^2-\lambda_{s}s_{i}+2\lambda_{b}s_{i}-\lambda_{b}\right\}\\
		&=(1-\varepsilon) \left(\lambda_{s}+\lambda_{b}\right).
	\end{aligned}
\end{equation}
Paired with the boundary above, there is $ s_i=0 $ as the end of $\{1,0,...\} $ part with $ s_{i-1}=1 $. For this
	\begin{equation}
	\begin{aligned}
		&\mathrm{E}\left\{\operatorname{Poi}\left(\varepsilon\left(\lambda_{s} s_{i-1}+\lambda_{b}\right)\right) \times\left(2 s_{i}-1\right)\right\}  \\
		&=\varepsilon\mathrm{E}\left\{2\lambda_{s} s_{i-1}s_{i}-\lambda_{s} s_{i-1}+2\lambda_{b}s_{i}-\lambda_{b}\right\}\\
		&=-\varepsilon( \lambda_{s}+\lambda_{b}),
	\end{aligned}
\end{equation}

\begin{equation}
	\begin{aligned}
		&\mathrm{E}\left\{\operatorname{Poi}\left(\left(1-\varepsilon\right)\left(\lambda_{s} s_{i}+\lambda_{b}\right)\right) \times\left(2 s_{i}-1\right)\right\} \\
		&=\left(1-\varepsilon\right)\mathrm{E}\left\{2\lambda_{s}s_{i}^2-\lambda_{s}s_{i}+2\lambda_{b}s_{i}-\lambda_{b}\right\}\\
		&=-(1-\varepsilon) \lambda_{b}.
	\end{aligned}
\end{equation}

Thus, the expectation of $ \operatorname{C}_{0,0}  $ is given by
	\begin{equation}
		\begin{aligned}
			&\mathrm{E}\left(\operatorname{C}_{0,0}\right)\\
			&=L(1-\alpha)\left[0+(1-\varepsilon) \frac{\lambda_{s}}{2}\right]+(\alpha L -2)\left[(-\varepsilon) \frac{\lambda_{s}}{2}+(1-\varepsilon) \frac{\lambda_{s}}{2}\right]\\
			&+\left[\varepsilon( \frac{\lambda_{s}}{2}+\lambda_{b})+(1-\varepsilon) \left(\lambda_{s}+\lambda_{b}\right)\right]+\left[-\varepsilon( \lambda_{s}+\lambda_{b})-(1-\varepsilon) \lambda_{b}\right]\\
			&=L \frac{\lambda_{s}}{2}[1-(1+\alpha) \varepsilon]+\varepsilon \frac{\lambda_{s}}{2}.
		\end{aligned}
	\end{equation}

Since two parts $ \operatorname{Poi}\left(\varepsilon\left(\lambda_{s} s_{i-1}+\lambda_{b}\right)\right) $ and from $ \operatorname{Poi}\left(\left(1-\varepsilon\right)\left(\lambda_{s} s_{i}+\lambda_{b}\right)\right) $ are statistically independent, whether $ s_i \in RANDOM~part ~or ~\{1,0,...\} ~part $, we have
\begin{equation}
	\begin{aligned}
		&\mathrm{E}\left\{\operatorname { V a r } \left(\operatorname{Poi}\left((1-\varepsilon)\left(\lambda_{s} s_{i}+\lambda_{b}\right)\right)\times\left(2 s_{i}-1\right)\right)\right\}\\
		&=\mathrm{E}\left\{\mathrm{E}\left[\operatorname { V a r } \left(\operatorname{Poi}\left((1-\varepsilon)\left(\lambda_{s} s_{i}+\lambda_{b}\right)\right)\times\left(2 s_{i}-1\right)\right)\mid \bm{s}\right]\right\}\\
		&=\mathrm{E}\left\{\mathrm{E}\left[ (1-\varepsilon)\left(\lambda_{s} s_{i}+\lambda_{b}\right)\times\left(2 s_{i}-1\right)^2\mid \bm{s}\right]\right\}\\
		&=\mathrm{E}\left\{(1-\varepsilon)\left(\lambda_{s} s_{i}+\lambda_{b}\right)\right\}\\
		&=(1-\varepsilon)\left(\frac{\lambda_{s}}{2}+\lambda_{b}\right), \\
	\end{aligned}
\end{equation}

\begin{equation}
	\begin{aligned}
		&\mathrm{E}\left\{\operatorname{Var}\left(\operatorname{Poi}\left(\varepsilon\left(\lambda_{s} s_{i-1}+\lambda_{b}\right)\right) \times\left(2 s_{i}-1\right)\right)\right\}\\
		&=\mathrm{E}\left\{\mathrm{E}\left[\operatorname { V a r } \left(\operatorname{Poi}\left(\varepsilon\left(\lambda_{s} s_{i-1}+\lambda_{b}\right)\right)\times\left(2 s_{i}-1\right)\right)\mid \bm{s}\right]\right\}\\
		&=\mathrm{E}\left\{\mathrm{E}\left[ \varepsilon\left(\lambda_{s} s_{i-1}+\lambda_{b}\right)\times\left(2 s_{i}-1\right)^2\mid \bm{s}\right]\right\}\\
		&=\mathrm{E}\left\{\varepsilon\left(\lambda_{s} s_{i-1}+\lambda_{b}\right)\right\}\\
		&=\varepsilon\left(\frac{\lambda_{s}}{2}+\lambda_{b}\right).
	\end{aligned}
\end{equation}

Thus, the expectation of variance of $ \operatorname{C}_{0,0}  $ is given by 
\begin{align}
	\mathrm{E}\left(\operatorname{Var}\left(\operatorname{C}_{0,0}\right)\right)=L\left(\frac{\lambda_{s}}{2}+\lambda_{b}\right)\left(1-\varepsilon+\varepsilon\right)=L\left(\frac{\lambda_{s}}{2}+\lambda_{b}\right).
\end{align}

\section{ Proof of Theorem \ref{theorem.meanvarforCase2} }  \label{appendix.meanvarforCase2}
Combining Equation \eqref{equ.correlation for case 2}, we have

\begin{equation}
	\operatorname{C}_{0,k}  \sim \sum_{i=1}^{L}\left\{n_{i} \times\left(2 s_{i}-1\right)\right\}
\end{equation}	
where
\begin{equation}
	n_{i} \sim \left[\operatorname{Poi}\left((1-\frac{k}{n})\left(\lambda_{s} s_{i}+\lambda_{b}\right)\right)+\operatorname{Poi}\left(\varepsilon\left(\lambda_{s} s_{i}+\lambda_{b}\right)\right)+\operatorname{Poi}\left((\frac{k}{n}-\varepsilon)\left(\lambda_{s} s_{i+1}+\lambda_{b}\right)\right)\right]
\end{equation} 

Firstly, if $ s_i$ lies in RANDOM part, we have $ \mathrm{E}\left(s_{i-1}s_{i}\right) = \frac{1}{4}$ when its two neighbors also lie in RANDOM part, and thus
\begin{equation}
	\begin{aligned}
		&\mathrm{E}\left\{\operatorname{Poi}\left(\left(1-\frac{k}{n}\right)\left(\lambda_{s} s_{i}+\lambda_{b}\right)\right) \times\left(2 s_{i}-1\right)\right\} \\
		&=\mathrm{E}\left\{\mathrm{E}\left[\operatorname{Poi}\left(\left(1-\frac{k}{n}\right)\left(\lambda_{s} s_{i}+\lambda_{b}\right)\right) \times\left(2 s_{i}-1\right)\mid \bm{s}\right]\right\} \\
		&=(1-\frac{k}{n})\mathrm{E}\left\{2\lambda_{s}s_{i}^2-\lambda_{s}s_{i}+2\lambda_{b}s_{i}-\lambda_{b}\right\}\\
		&=(1-\frac{k}{n}) \frac{\lambda_{s}}{2},
	\end{aligned}
\end{equation}

\begin{equation}
	\begin{aligned}
		&\mathrm{E}\left\{\operatorname{Poi}\left(\varepsilon\left(\lambda_{s} s_{i}+\lambda_{b}\right)\right) \times\left(2 s_{i}-1\right)\right\} \\
		&=\mathrm{E}\left\{\mathrm{E}\left[\operatorname{Poi}\left(\varepsilon\left(\lambda_{s} s_{i}+\lambda_{b}\right)\right) \times\left(2 s_{i}-1\right)\mid \bm{s}\right]\right\} \\
		&=\varepsilon\mathrm{E}\left\{2\lambda_{s}s_{i}^2-\lambda_{s}s_{i}+2\lambda_{b}s_{i}-\lambda_{b}\right\}\\
		&=\varepsilon \frac{\lambda_{s}}{2},
	\end{aligned}
\end{equation}

\begin{equation}
	\begin{aligned}
		&\mathrm{E}\left\{\operatorname{Poi}\left(\left(\frac{k}{n}-\varepsilon\right)\left(\lambda_{s} s_{i+1}+\lambda_{b}\right)\right) \times\left(2 s_{i}-1\right)\right\} \\
		&=\mathrm{E}\left\{\mathrm{E}\left[\operatorname{Poi}\left(\left(\frac{k}{n}-\varepsilon\right)\left(\lambda_{s} s_{i+1}+\lambda_{b}\right)\right) \times\left(2 s_{i}-1\right)\mid \bm{s}\right]\right\} \\
		&=\left(\frac{k}{n}-\varepsilon\right)\mathrm{E}\left\{2\lambda_{s}s_{i}s_{i+1}-\lambda_{s}s_{i+1}+2\lambda_{b}s_{i}-\lambda_{b}\right\}\\
		&=0.
	\end{aligned}
\end{equation}
If $ s_i$ lies in RANDOM part, we have $ s_{i+1}=1$ when the boundary is between  $ s_{i}$ and  $ s_{i+1}$ or we have $ s_{i+1}=0$ when $ i=L $, and thus
\begin{equation}
	\begin{aligned}
		&\mathrm{E}\left\{\operatorname{Poi}\left(\left(1-\frac{k}{n}\right)\left(\lambda_{s} s_{i}+\lambda_{b}\right)\right) \times\left(2 s_{i}-1\right)\right\} \\
		&=(1-\frac{k}{n})\mathrm{E}\left\{2\lambda_{s}s_{i}^2-\lambda_{s}s_{i}+2\lambda_{b}s_{i}-\lambda_{b}\right\}\\
		&=(1-\frac{k}{n}) \frac{\lambda_{s}}{2},
	\end{aligned}
\end{equation}

\begin{equation}
	\begin{aligned}
		&\mathrm{E}\left\{\operatorname{Poi}\left(\varepsilon\left(\lambda_{s} s_{i}+\lambda_{b}\right)\right) \times\left(2 s_{i}-1\right)\right\} \\
		&=\varepsilon\mathrm{E}\left\{2\lambda_{s}s_{i}^2-\lambda_{s}s_{i}+2\lambda_{b}s_{i}-\lambda_{b}\right\}\\
		&=\varepsilon \frac{\lambda_{s}}{2},
	\end{aligned}
\end{equation}
whether $ s_{i+1} =0 $ or $ s_{i+1}=1 $, we have
\begin{equation}
	\begin{aligned}
		&\mathrm{E}\left\{\operatorname{Poi}\left(\left(\frac{k}{n}-\varepsilon\right)\left(\lambda_{s} s_{i+1}+\lambda_{b}\right)\right) \times\left(2 s_{i}-1\right)\right\} \\
		&=\left(\frac{k}{n}-\varepsilon\right)\mathrm{E}\left\{2\lambda_{s}s_{i}s_{i+1}-\lambda_{s}s_{i+1}+2\lambda_{b}s_{i}-\lambda_{b}\right\}\\
		&=0.
	\end{aligned}
\end{equation}

If $ s_i $ lies in $\{1,0,...\} $ part, we have$ \mathrm{E}\left(s_{i+1}s_{i}\right) = 0$ when its two neighbors also lie in $\{1,0,...\} $ part.  For $ (\alpha L - 2) $ items which don't lie in boundaries, we have $ \mathrm{E}\left(s_{i}\right) =\frac{1}{2}$ and $ \mathrm{E}\left(s_{i+1}\right) =\frac{1}{2}$.
\begin{align} 
	\mathrm{E}\left\{\operatorname{Poi}\left(\left(1-\frac{k}{n}\right)\left(\lambda_{s} s_{i}+\lambda_{b}\right)\right) \times\left(2 s_{i}-1\right)\right\} &=(1-\frac{k}{n}) \frac{\lambda_{s}}{2},\\
	\mathrm{E}\left\{\operatorname{Poi}\left(\varepsilon\left(\lambda_{s} s_{i}+\lambda_{b}\right)\right) \times\left(2 s_{i}-1\right)\right\}  &=\varepsilon \frac{\lambda_{s}}{2},
\end{align}

\begin{equation}
	\begin{aligned}
		&\mathrm{E}\left\{\operatorname{Poi}\left(\left(\frac{k}{n}-\varepsilon\right)\left(\lambda_{s} s_{i+1}+\lambda_{b}\right)\right) \times\left(2 s_{i}-1\right)\right\} \\
		&=\mathrm{E}\left\{\mathrm{E}\left[\operatorname{Poi}\left(\left(\frac{k}{n}-\varepsilon\right)\left(\lambda_{s} s_{i+1}+\lambda_{b}\right)\right) \times\left(2 s_{i}-1\right)\mid \bm{s}\right]\right\} \\
		&=\left(\frac{k}{n}-\varepsilon\right)\mathrm{E}\left\{2\lambda_{s}s_{i}s_{i+1}-\lambda_{s}s_{i+1}+2\lambda_{b}s_{i}-\lambda_{b}\right\}\\
		&=-(\frac{k}{n}-\varepsilon) \frac{\lambda_{s}}{2}.
	\end{aligned}
\end{equation}
If $ s_i$ lies in $\{1,0,...\} $ part, we have $ s_{i+1}$ lies in RANDOM part and $ s_i=0 $ when the boundary is between  $ s_{i}$ and $ s_{i+1}$, and thus
\begin{equation}
	\begin{aligned}
		&\mathrm{E}\left\{\operatorname{Poi}\left(\left(1-\frac{k}{n}\right)\left(\lambda_{s} s_{i}+\lambda_{b}\right)\right) \times\left(2 s_{i}-1\right)\right\} \\
		&=(1-\frac{k}{n})\mathrm{E}\left\{2\lambda_{s}s_{i}^2-\lambda_{s}s_{i}+2\lambda_{b}s_{i}-\lambda_{b}\right\}\\
		&=(1-\frac{k}{n}) \left(-\lambda_{b}\right),
	\end{aligned}
\end{equation}

\begin{equation}
	\begin{aligned}
		&\mathrm{E}\left\{\operatorname{Poi}\left(\varepsilon\left(\lambda_{s} s_{i}+\lambda_{b}\right)\right) \times\left(2 s_{i}-1\right)\right\} \\
		&=\varepsilon\mathrm{E}\left\{2\lambda_{s}s_{i}^2-\lambda_{s}s_{i}+2\lambda_{b}s_{i}-\lambda_{b}\right\}\\
		&=\varepsilon \left(-\lambda_{b}\right),
	\end{aligned}
\end{equation}


\begin{equation}
	\begin{aligned}
		&\mathrm{E}\left\{\operatorname{Poi}\left(\left(\frac{k}{n}-\varepsilon\right)\left(\lambda_{s} s_{i+1}+\lambda_{b}\right)\right) \times\left(2 s_{i}-1\right)\right\} \\
		&=\left(\frac{k}{n}-\varepsilon\right)\mathrm{E}\left\{2\lambda_{s}s_{i}s_{i+1}-\lambda_{s}s_{i+1}+2\lambda_{b}s_{i}-\lambda_{b}\right\}\\
		&=-(\frac{k}{n}-\varepsilon)(\frac{\lambda_{s}}{2}+\lambda_{b}).
	\end{aligned}
\end{equation}

Paired with the boundary above, there is $ s_i=1 $ as the start of $\{1,0,...\} $ part with $ s_{i+1}=0 $. For this
\begin{equation}
	\begin{aligned}
		&\mathrm{E}\left\{\operatorname{Poi}\left(\left(1-\frac{k}{n}\right)\left(\lambda_{s} s_{i}+\lambda_{b}\right)\right) \times\left(2 s_{i}-1\right)\right\} \\
		&=(1-\frac{k}{n})\mathrm{E}\left\{2\lambda_{s}s_{i}^2-\lambda_{s}s_{i}+2\lambda_{b}s_{i}-\lambda_{b}\right\}\\
		&=(1-\frac{k}{n}) \left(\lambda_{s}+\lambda_{b}\right),
	\end{aligned}
\end{equation}

\begin{equation}
	\begin{aligned}
		&\mathrm{E}\left\{\operatorname{Poi}\left(\varepsilon\left(\lambda_{s} s_{i}+\lambda_{b}\right)\right) \times\left(2 s_{i}-1\right)\right\} \\
		&=\varepsilon\mathrm{E}\left\{2\lambda_{s}s_{i}^2-\lambda_{s}s_{i}+2\lambda_{b}s_{i}-\lambda_{b}\right\}\\
		&=\varepsilon \left(\lambda_{s}+\lambda_{b}\right),
	\end{aligned}
\end{equation}

\begin{equation}
	\begin{aligned}
		&\mathrm{E}\left\{\operatorname{Poi}\left(\left(\frac{k}{n}-\varepsilon\right)\left(\lambda_{s} s_{i+1}+\lambda_{b}\right)\right) \times\left(2 s_{i}-1\right)\right\} \\
		&=\left(\frac{k}{n}-\varepsilon\right)\mathrm{E}\left\{2\lambda_{s}s_{i}s_{i+1}-\lambda_{s}s_{i+1}+2\lambda_{b}s_{i}-\lambda_{b}\right\}\\
		&=(\frac{k}{n}-\varepsilon)(\lambda_{b}).
	\end{aligned}
\end{equation}

Thus, the expectation of $ \operatorname{C}_{0,k}  $ is given by
\begin{align} 
	\mathrm{E}\left(\operatorname{C}_{0,k}\right)&=L \frac{\lambda_{s}}{2}[(1-\alpha)(1-\frac{k}{n}+\varepsilon)+\alpha(1-\frac{k}{n}+\varepsilon-(\frac{k}{n}-\varepsilon))]+(\frac{k}{n}-\varepsilon)\frac{\lambda_{s}}{2}\\
	&=L \frac{\lambda_{s}}{2}[1-(1+\alpha) (\frac{k}{n}-\varepsilon)]+(\frac{k}{n}-\varepsilon)\frac{\lambda_{s}}{2}.
\end{align}

Similar to that of Case 1, the expectation of variance of $ \operatorname{Corr}_{0,k}  $ is given by
\begin{align}
	\mathrm{E}\left(\operatorname{Var}\left(\operatorname{C}_{0,k}\right)\right)=L\left(\frac{\lambda_{s}}{2}+\lambda_{b}\right).
\end{align}

\section{ Proof of Theorem \ref{theorem.covforCase2} }  \label{appendix.covforCase2}
\begin{equation} 
	\operatorname{C}_{0,0}  \sim \sum_{i=1}^{L}\left\{n_{i} \times\left(2 s_{i}-1\right)\right\},
\end{equation}
where
\begin{equation}
	n_{i} \sim \left[\operatorname{Poi}\left(\varepsilon\left(\lambda_{s} s_{i-1}+\lambda_{b}\right)\right)+\operatorname{Poi}\left((\frac{k}{n}-\varepsilon)\left(\lambda_{s} s_{i}+\lambda_{b}\right)\right)+\operatorname{Poi}\left((1-\frac{k}{n})\left(\lambda_{s} s_{i}+\lambda_{b}\right)\right)\right].
\end{equation} 

By analysis, the covariance is mainly composed of covariance of Poisson random variables, such as $ \operatorname{Cov}\left\{\left(2 s_{i}-1\right) \operatorname{Poi}\left(\varepsilon\left(\lambda_{s} s_{i-1}+\lambda_{b}\right)\right),\left(2 s_{j}-1\right) \operatorname{Poi}\left(\varepsilon\left(\lambda_{s} s_{j}+\lambda_{b}\right)\right)\right\} $ when $ i-1=j $, which means $\operatorname{Poi}\left(\varepsilon\left(\lambda_{s} s_{i-1}+\lambda_{b}\right)\right)$ and $ \operatorname{Poi}\left(\varepsilon\left(\lambda_{s} s_{j}+\lambda_{b}\right)\right) $ are actually the same random variable as a result that they are from Poisson process during the same period of time.

Firstly, $ s_j$ lies in RANDOM part. 

For $ \left(1-\alpha\right)L $ pairs of $ \left(i,j\right) $ with $ i=j $, we have
\begin{equation}
	\begin{aligned}
		&\mathrm{E}\left(\operatorname{Cov}\left\{\left(2 s_{i}-1\right) \operatorname{Poi}\left(\left(1-\frac{k}{n}\right)\left(\lambda_{s} s_{i}+\lambda_{b}\right)\right),\left(2 s_{j}-1\right) \operatorname{Poi}\left(\left(1-\frac{k}{n}\right)\left(\lambda_{s} s_{j}+\lambda_{b}\right)\right)\right\}\right)\\
		&=\mathrm{E}\left\{\mathrm{E}\left(\operatorname{Cov}\left\{\left(2 s_{i}-1\right) \operatorname{Poi}\left(\left(1-\frac{k}{n}\right)\left(\lambda_{s} s_{i}+\lambda_{b}\right)\right),\left(2 s_{j}-1\right) \operatorname{Poi}\left(\left(1-\frac{k}{n}\right)\left(\lambda_{s} s_{j}+\lambda_{b}\right)\right)\right\}\right)\mid \bm{s} \right\}  \\
		&=\mathrm{E}\left\{\left(2 s_{j}-1\right)\left(2 s_{j}-1\right)\left(\left(1-\frac{k}{n}\right)\left(\lambda_{s} s_{j}+\lambda_{b}\right)\right)\right\}\\
		&=\left(1-\frac{k}{n}\right)\left(\frac{\lambda_{s}}{2}+\lambda_{b}\right).
	\end{aligned}
\end{equation}

If $ s_j$ lies in RANDOM part, we have $ \mathrm{E}\left(s_{j}s_{j+1}\right) = \frac{1}{4}$ when its two neighbors also lie in RANDOM part.

For $ \left[\left(1-\alpha\right)L-2\right] $ pairs of $ \left(i,j\right) $ with $ i-1=j $, 
\begin{equation}
	\begin{aligned}
		&\mathrm{E}\left(\operatorname{Cov}\left\{\left(2 s_{i}-1\right) \operatorname{Poi}\left(\varepsilon\left(\lambda_{s} s_{i-1}+\lambda_{b}\right)\right),\left(2 s_{j}-1\right) \operatorname{Poi}\left(\varepsilon\left(\lambda_{s} s_{j}+\lambda_{b}\right)\right)\right\}\right)\\
		&=\mathrm{E}\left\{\mathrm{E}\left(\operatorname{Cov}\left\{\left(2 s_{j+1}-1\right) \operatorname{Poi}\left(\varepsilon\left(\lambda_{s} s_{j}+\lambda_{b}\right)\right),\left(2 s_{j}-1\right) \operatorname{Poi}\left(\varepsilon\left(\lambda_{s} s_{j}+\lambda_{b}\right)\right)\right\}\right)\mid \bm{s} \right\}  \\
		&=\mathrm{E}\left\{\left(2 s_{j+1}-1\right)\left(2 s_{j}-1\right)\left(\varepsilon\left(\lambda_{s} s_{j}+\lambda_{b}\right)\right)\right\}\\
		&=0,\\
	\end{aligned}
\end{equation}

\begin{equation}
	\begin{aligned}
		&\mathrm{E}\left(\operatorname{Cov}\left\{\left(2 s_{i}-1\right) \operatorname{Poi}\left(\left(\frac{k}{n}-\varepsilon\right)\left(\lambda_{s} s_{i}+\lambda_{b}\right)\right),\left(2 s_{j}-1\right) \operatorname{Poi}\left(\left(\frac{k}{n}-\varepsilon\right)\left(\lambda_{s} s_{j+1}+\lambda_{b}\right)\right)\right\}\right)\\
		&=\mathrm{E}\left\{\mathrm{E}\left(\operatorname{Cov}\left\{\left(2 s_{i}-1\right) \operatorname{Poi}\left(\left(\frac{k}{n}-\varepsilon\right)\left(\lambda_{s} s_{i}+\lambda_{b}\right)\right),\left(2 s_{j}-1\right) \operatorname{Poi}\left(\left(\frac{k}{n}-\varepsilon\right)\left(\lambda_{s} s_{j+1}+\lambda_{b}\right)\right)\right\}\right)\mid \bm{s} \right\}  \\
		&=\mathrm{E}\left\{\left(2 s_{j+1}-1\right)\left(2 s_{j}-1\right)\left(\left(\frac{k}{n}-\varepsilon\right)\left(\lambda_{s} s_{j+1}+\lambda_{b}\right)\right)\right\}\\
		&=0.\\
	\end{aligned}
\end{equation}

If $ s_j$ lies in RANDOM part, we have $ s_{j+1}=1$ when the boundary is between  $ s_{j}$ and  $ s_{j+1}$ and we have $ s_{j+1}=0 $ when $ j=L $.

For $ 2 $ pairs of $ \left(i,j\right) $ with $ i-1=j $, 
\begin{equation}
	\begin{aligned}
		&\mathrm{E}\left(\operatorname{Cov}\left\{\left(2 s_{i}-1\right) \operatorname{Poi}\left(\varepsilon\left(\lambda_{s} s_{i-1}+\lambda_{b}\right)\right),\left(2 s_{j}-1\right) \operatorname{Poi}\left(\varepsilon\left(\lambda_{s} s_{j}+\lambda_{b}\right)\right)\right\}\right)\\
		&=\mathrm{E}\left\{\mathrm{E}\left(\operatorname{Cov}\left\{\left(2 s_{i}-1\right) \operatorname{Poi}\left(\varepsilon\left(\lambda_{s} s_{i-1}+\lambda_{b}\right)\right),\left(2 s_{j}-1\right) \operatorname{Poi}\left(\varepsilon\left(\lambda_{s} s_{j}+\lambda_{b}\right)\right)\right\}\right)\mid \bm{s} \right\}  \\
		&=\mathrm{E}\left\{\left(2 s_{j+1}-1\right)\left(2 s_{j}-1\right)\left(\varepsilon\left(\lambda_{s} s_{j}+\lambda_{b}\right)\right)\right\}\\
		&=\varepsilon\mathrm{E}\left(4\lambda_s s_{j}^2 s_{j+1} -2\lambda_{s} s_{j}^2-2 \lambda_{s} s_{j} s_{j+1} +\lambda_{s}s_{j} +4\lambda_{b} s_{j} s_{j+1} -2 \lambda_{b} s_{j}-2\lambda_{b} s_{j+1} +\lambda_{b}\right)\\
		&=0,\\
	\end{aligned}
\end{equation}

\begin{equation}
	\begin{aligned}
		&\mathrm{E}\left(\operatorname{Cov}\left\{\left(2 s_{i}-1\right) \operatorname{Poi}\left(\left(\frac{k}{n}-\varepsilon\right)\left(\lambda_{s} s_{i}+\lambda_{b}\right)\right),\left(2 s_{j}-1\right) \operatorname{Poi}\left(\left(\frac{k}{n}-\varepsilon\right)\left(\lambda_{s} s_{j+1}+\lambda_{b}\right)\right)\right\}\right)\\
		&=\mathrm{E}\left\{\mathrm{E}\left(\operatorname{Cov}\left\{\left(2 s_{i}-1\right) \operatorname{Poi}\left(\left(\frac{k}{n}-\varepsilon\right)\left(\lambda_{s} s_{i}+\lambda_{b}\right)\right),\left(2 s_{j}-1\right) \operatorname{Poi}\left(\left(\frac{k}{n}-\varepsilon\right)\left(\lambda_{s} s_{j+1}+\lambda_{b}\right)\right)\right\}\right)\mid \bm{s} \right\}  \\
		&=\mathrm{E}\left\{\left(2 s_{j+1}-1\right)\left(2 s_{j}-1\right)\left(\left(\frac{k}{n}-\varepsilon\right)\left(\lambda_{s} s_{j+1}+\lambda_{b}\right)\right)\right\}\\
		&=\left(\frac{k}{n}-\varepsilon\right)\mathrm{E}\left(4\lambda_s s_{j} s_{j+1}^2 -2\lambda_{s} s_{j} s_{j+1}-2 \lambda_{s} s_{j+1}^2 +\lambda_{s}s_{j+1} +4\lambda_{b} s_{j} s_{j+1} -2 \lambda_{b} s_{j}-2\lambda_{b} s_{j+1} +\lambda_{b}\right)\\
		&=0.\\
	\end{aligned}
\end{equation}


On the other hand, $ s_j$ lies in $\{1,0,...\} $ part. 

For $ \alpha L $ pairs of $ \left(i,j\right) $ with $ i=j $,
\begin{equation}
	\begin{aligned}
		&\mathrm{E}\left(\operatorname{Cov}\left\{\left(2 s_{i}-1\right) \operatorname{Poi}\left(\left(1-\frac{k}{n}\right)\left(\lambda_{s} s_{i}+\lambda_{b}\right)\right),\left(2 s_{j}-1\right) \operatorname{Poi}\left(\left(1-\frac{k}{n}\right)\left(\lambda_{s} s_{j}+\lambda_{b}\right)\right)\right\}\right)\\
		&=\mathrm{E}\left\{\mathrm{E}\left(\operatorname{Cov}\left\{\left(2 s_{i}-1\right) \operatorname{Poi}\left(\left(1-\frac{k}{n}\right)\left(\lambda_{s} s_{i}+\lambda_{b}\right)\right),\left(2 s_{j}-1\right) \operatorname{Poi}\left(\left(1-\frac{k}{n}\right)\left(\lambda_{s} s_{j}+\lambda_{b}\right)\right)\right\}\right)\mid \bm{s} \right\}  \\
		&=\mathrm{E}\left\{\left(2 s_{j}-1\right)\left(2 s_{j}-1\right)\left(\left(1-\frac{k}{n}\right)\left(\lambda_{s} s_{j}+\lambda_{b}\right)\right)\right\}\\
		&=\left(1-\frac{k}{n}\right)\left(\frac{\lambda_{s}}{2}+\lambda_{b}\right).
	\end{aligned}
\end{equation}

If $ s_j $ lies in $\{1,0,...\} $ part, we have $ \mathrm{E}\left(s_{j}s_{j+1}\right) = 0$ when its two neighbors also lie in $\{1,0,...\} $ part. There are $ (\alpha L-2) $ items not lying around boundaries with $ \mathrm{E}\left(s_{j}\right) =\frac{1}{2}$ and $ \mathrm{E}\left(s_{j+1}\right) =\frac{1}{2}$. For $ (\alpha L-2) $ pairs of $ \left(i,j\right) $ with $ i-1=j $, we have

\begin{equation}
	\begin{aligned}
		&\mathrm{E}\left(\operatorname{Cov}\left\{\left(2 s_{i}-1\right) \operatorname{Poi}\left(\varepsilon\left(\lambda_{s} s_{i-1}+\lambda_{b}\right)\right),\left(2 s_{j}-1\right) \operatorname{Poi}\left(\varepsilon\left(\lambda_{s} s_{j}+\lambda_{b}\right)\right)\right\}\right)\\
		&=\mathrm{E}\left\{\mathrm{E}\left(\operatorname{Cov}\left\{\left(2 s_{i}-1\right) \operatorname{Poi}\left(\varepsilon\left(\lambda_{s} s_{i-1}+\lambda_{b}\right)\right),\left(2 s_{j}-1\right) \operatorname{Poi}\left(\varepsilon\left(\lambda_{s} s_{j}+\lambda_{b}\right)\right)\right\}\right)\mid \bm{s} \right\}  \\
		&=\mathrm{E}\left\{\left(2 s_{j}-1\right)\left(2 s_{j+1}-1\right)\left(\varepsilon\left(\lambda_{s} s_{j}+\lambda_{b}\right)\right)\right\}\\
		&=\left(- \varepsilon \right)\left(\frac{\lambda_{s}}{2}+\lambda_{b}\right),
	\end{aligned}
\end{equation}

\begin{equation}
	\begin{aligned}
		&\mathrm{E}\left(\operatorname{Cov}\left\{\left(2 s_{i}-1\right) \operatorname{Poi}\left(\left(\frac{k}{n}-\varepsilon\right)\left(\lambda_{s} s_{i}+\lambda_{b}\right)\right),\left(2 s_{j}-1\right) \operatorname{Poi}\left(\left(\frac{k}{n}-\varepsilon\right)\left(\lambda_{s} s_{j+1}+\lambda_{b}\right)\right)\right\}\right)\\
		&=\mathrm{E}\left\{\mathrm{E}\left(\operatorname{Cov}\left\{\left(2 s_{i}-1\right) \operatorname{Poi}\left(\left(\frac{k}{n}-\varepsilon\right)\left(\lambda_{s} s_{i}+\lambda_{b}\right)\right),\left(2 s_{j}-1\right) \operatorname{Poi}\left(\left(\frac{k}{n}-\varepsilon\right)\left(\lambda_{s} s_{j+1}+\lambda_{b}\right)\right)\right\}\right)\mid \bm{s} \right\}  \\
		&=\mathrm{E}\left\{\left(2 s_{j}-1\right)\left(2 s_{j+1}-1\right)\left(\left(\frac{k}{n}-\varepsilon\right)\left(\lambda_{s} s_{j+1}+\lambda_{b}\right)\right)\right\}\\
		&= -\left(\frac{k}{n}-\varepsilon \right)\left(\frac{\lambda_{s}}{2}+\lambda_{b}\right).
	\end{aligned}
\end{equation}

If $ s_j$ lies in $\{1,0,...\} $ part, we have $ s_{j+1}$ lies in RANDOM part and $ s_j=0 $ when the boundary is between  $ s_{j}$ and $ s_{j+1}$. For 1 pair of $ \left(i,j\right) $ with $ i-1=j $, we have
\begin{equation}
	\begin{aligned}
		&\mathrm{E}\left(\operatorname{Cov}\left\{\left(2 s_{i}-1\right) \operatorname{Poi}\left(\varepsilon\left(\lambda_{s} s_{i-1}+\lambda_{b}\right)\right),\left(2 s_{j}-1\right) \operatorname{Poi}\left(\varepsilon\left(\lambda_{s} s_{j}+\lambda_{b}\right)\right)\right\}\right)\\
		&=\mathrm{E}\left\{\mathrm{E}\left(\operatorname{Cov}\left\{\left(2 s_{i}-1\right) \operatorname{Poi}\left(\varepsilon\left(\lambda_{s} s_{i-1}+\lambda_{b}\right)\right),\left(2 s_{j}-1\right) \operatorname{Poi}\left(\varepsilon\left(\lambda_{s} s_{j}+\lambda_{b}\right)\right)\right\}\right)\mid \bm{s} \right\}  \\
		&=\mathrm{E}\left\{\left(2 s_{j}-1\right)\left(2 s_{j+1}-1\right)\left(\varepsilon\left(\lambda_{s} s_{j}+\lambda_{b}\right)\right)\right\}\\
		&=\varepsilon\mathrm{E}\left(4\lambda_s s_{j}^2 s_{j+1} -2\lambda_{s} s_{j}^2-2 \lambda_{s} s_{j} s_{j+1} +\lambda_{s}s_{j} +4\lambda_{b} s_{j} s_{j+1} -2 \lambda_{b} s_{j}-2\lambda_{b} s_{j+1} +\lambda_{b}\right)\\
		&=0,
	\end{aligned}
\end{equation}

\begin{equation}
	\begin{aligned}
		&\mathrm{E}\left(\operatorname{Cov}\left\{\left(2 s_{i}-1\right) \operatorname{Poi}\left(\left(\frac{k}{n}-\varepsilon\right)\left(\lambda_{s} s_{i}+\lambda_{b}\right)\right),\left(2 s_{j}-1\right) \operatorname{Poi}\left(\left(\frac{k}{n}-\varepsilon\right)\left(\lambda_{s} s_{j+1}+\lambda_{b}\right)\right)\right\}\right)\\
		&=\mathrm{E}\left\{\mathrm{E}\left(\operatorname{Cov}\left\{\left(2 s_{i}-1\right) \operatorname{Poi}\left(\left(\frac{k}{n}-\varepsilon\right)\left(\lambda_{s} s_{i}+\lambda_{b}\right)\right),\left(2 s_{j}-1\right) \operatorname{Poi}\left(\left(\frac{k}{n}-\varepsilon\right)\left(\lambda_{s} s_{j+1}+\lambda_{b}\right)\right)\right\}\right)\mid \bm{s} \right\}  \\
		&=\mathrm{E}\left\{\left(2 s_{j}-1\right)\left(2 s_{j+1}-1\right)\left(\left(\frac{k}{n}-\varepsilon\right)\left(\lambda_{s} s_{j+1}+\lambda_{b}\right)\right)\right\}\\
		&=\left(\frac{k}{n}-\varepsilon\right)\mathrm{E}\left(4\lambda_s s_{j} s_{j+1}^2 -2\lambda_{s} s_{j} s_{j+1}-2 \lambda_{s} s_{j+1}^2 +\lambda_{s}s_{j+1} +4\lambda_{b} s_{j} s_{j+1} -2 \lambda_{b} s_{j}-2\lambda_{b} s_{j+1} +\lambda_{b}\right)\\
		&= -\left(\frac{k}{n}-\varepsilon\right)\frac{\lambda_{s}}{2}.
	\end{aligned}
\end{equation}

Paired with the boundary above, there is $ s_j=1 $ as the start of $\{1,0,...\} $ part with $ s_{j+1}=0 $. For 1 pair of $ \left(i,j\right) $ with $ i-1=j $, we have
\begin{equation}
	\begin{aligned}
		&\mathrm{E}\left(\operatorname{Cov}\left\{\left(2 s_{i}-1\right) \operatorname{Poi}\left(\varepsilon\left(\lambda_{s} s_{i-1}+\lambda_{b}\right)\right),\left(2 s_{j}-1\right) \operatorname{Poi}\left(\varepsilon\left(\lambda_{s} s_{j}+\lambda_{b}\right)\right)\right\}\right)\\
		&=\mathrm{E}\left\{\mathrm{E}\left(\operatorname{Cov}\left\{\left(2 s_{i}-1\right) \operatorname{Poi}\left(\varepsilon\left(\lambda_{s} s_{i-1}+\lambda_{b}\right)\right),\left(2 s_{j}-1\right) \operatorname{Poi}\left(\varepsilon\left(\lambda_{s} s_{j}+\lambda_{b}\right)\right)\right\}\right)\mid \bm{s} \right\}  \\
		&=\mathrm{E}\left\{\left(2 s_{j}-1\right)\left(2 s_{j+1}-1\right)\left(\varepsilon\left(\lambda_{s} s_{j}+\lambda_{b}\right)\right)\right\}\\
		&=\varepsilon\mathrm{E}\left(4\lambda_s s_{j}^2 s_{j+1} -2\lambda_{s} s_{j}^2-2 \lambda_{s} s_{j} s_{j+1} +\lambda_{s}s_{j} +4\lambda_{b} s_{j} s_{j+1} -2 \lambda_{b} s_{j}-2\lambda_{b} s_{j+1} +\lambda_{b}\right)\\
		&=-\varepsilon \left(\lambda_{s}+\lambda_{b}\right),
	\end{aligned}
\end{equation}

\begin{equation}
	\begin{aligned}
		&\mathrm{E}\left(\operatorname{Cov}\left\{\left(2 s_{i}-1\right) \operatorname{Poi}\left(\left(\frac{k}{n}-\varepsilon\right)\left(\lambda_{s} s_{i}+\lambda_{b}\right)\right),\left(2 s_{j}-1\right) \operatorname{Poi}\left(\left(\frac{k}{n}-\varepsilon\right)\left(\lambda_{s} s_{j+1}+\lambda_{b}\right)\right)\right\}\right)\\
		&=\mathrm{E}\left\{\mathrm{E}\left(\operatorname{Cov}\left\{\left(2 s_{i}-1\right) \operatorname{Poi}\left(\left(\frac{k}{n}-\varepsilon\right)\left(\lambda_{s} s_{i}+\lambda_{b}\right)\right),\left(2 s_{j}-1\right) \operatorname{Poi}\left(\left(\frac{k}{n}-\varepsilon\right)\left(\lambda_{s} s_{j+1}+\lambda_{b}\right)\right)\right\}\right)\mid \bm{s} \right\}  \\
		&=\mathrm{E}\left\{\left(2 s_{j}-1\right)\left(2 s_{j+1}-1\right)\left(\left(\frac{k}{n}-\varepsilon\right)\left(\lambda_{s} s_{j+1}+\lambda_{b}\right)\right)\right\}\\
		&=\left(\frac{k}{n}-\varepsilon\right)\mathrm{E}\left(4\lambda_s s_{j} s_{j+1}^2 -2\lambda_{s} s_{j} s_{j+1}-2 \lambda_{s} s_{j+1}^2 +\lambda_{s}s_{j+1} +4\lambda_{b} s_{j} s_{j+1} -2 \lambda_{b} s_{j}-2\lambda_{b} s_{j+1} +\lambda_{b}\right)\\
		&= -\left(\frac{k}{n}-\varepsilon \right)\lambda_{b}.
	\end{aligned}
\end{equation}

By adding up components above, the mathematical expectation of covariance is,
\begin{align}
	\mathrm{E}\left\{\operatorname{Cov}\left(\operatorname{C}_{0,0}, \operatorname{C}_{0, k}\right)\right\}
	=L\left(\frac{\lambda_{s}}{2}+\lambda_{b}\right)\left[1-(1+\alpha) \frac{k}{n}\right]+\frac{k}{n}\left(\frac{\lambda_{s}}{2}+\lambda_{b}\right)-\varepsilon \frac{\lambda_{s}}{2}.
\end{align}

\section{ Proof of Theorem \ref{theorem.meanvarforCase3} }  \label{appendix.meanvarforCase3}
Combining Equation \eqref{equ.correlation for case 3}, we have the proof as follows.
\begin{equation} 
	\operatorname{C}_{2m,k}  \sim \sum_{i=1}^{L}\left\{n_{i} \times\left(2 s_{i}-1\right)\right\},~~2m \leq \min\{\alpha L, \frac{1-\alpha}{2}L\},
\end{equation}	
where
\begin{equation}
	n_{i} \sim \left[\operatorname{Poi}\left((1-\frac{k}{n})\left(\lambda_{s} s_{i+2m}+\lambda_{b}\right)\right)+\operatorname{Poi}\left(\varepsilon\left(\lambda_{s} s_{i+2m}+\lambda_{b}\right)\right)+\operatorname{Poi}\left((\frac{k}{n}-\varepsilon)\left(\lambda_{s} s_{i+2m+1}+\lambda_{b}\right)\right)\right]
\end{equation} 
First, $ s_i$ lies in RANDOM part. We have $ \mathrm{E}\left(s_{i}s_{i+2m}\right) = \frac{1}{4}$ and $ \mathrm{E}\left(s_{i}s_{i+2m+1}\right) = \frac{1}{4}$ when $1 \leq i \leq (L-2m-1) $. $ s_{j},s_{j+2m},s_{j+2m+1} $ lie in RANDOM part. There are $ \left(1-\alpha\right)L-2m-1 $ cumulative items in the $ \operatorname{C}_{2m,k} $ follows
\begin{equation}
	\mathrm{E}\left\{n_{i} \times\left(2 s_{i}-1\right)\right\}=0, 1 \leq i \leq (L-2m-1).
\end{equation}

We have $ s_{i+2m} $ lying in RANDOM part as well as $ s_{i+2m+1} = 0$ when $ i=L-2m $. There are $ 1 $ cumulative item in the $ \operatorname{C}_{2m,k} $ follows
\begin{equation}
	\begin{aligned}
		&\mathrm{E}\left\{\operatorname{Poi}\left(\left(1-\frac{k}{n}\right)\left(\lambda_{s} s_{i+2m}+\lambda_{b}\right)\right) \times\left(2 s_{i}-1\right)\right\} \\
		&=\mathrm{E}\left\{\mathrm{E}\left[\operatorname{Poi}\left(\left(1-\frac{k}{n}\right)\left(\lambda_{s} s_{i+2m}+\lambda_{b}\right)\right) \times\left(2 s_{i}-1\right)\mid \bm{s}\right]\right\} \\
		&=(1-\frac{k}{n})\mathrm{E}\left\{2\lambda_{s}s_{i}s_{i+2m}-\lambda_{s}s_{i+2m}+2\lambda_{b}s_{i}-\lambda_{b}\right\}\\
		&=0,
	\end{aligned}
\end{equation}

\begin{equation}
	\begin{aligned}
		&\mathrm{E}\left\{\operatorname{Poi}\left(\varepsilon\left(\lambda_{s} s_{i+2m}+\lambda_{b}\right)\right) \times\left(2 s_{i}-1\right)\right\}  \\
		&=\mathrm{E}\left\{\mathrm{E}\left[\operatorname{Poi}\left(\varepsilon\left(\lambda_{s} s_{i+2m}+\lambda_{b}\right)\right) \times\left(2 s_{i}-1\right)\mid \bm{s}\right]\right\} \\
		&=\varepsilon\mathrm{E}\left\{2\lambda_{s}s_{i}s_{i+2m}-\lambda_{s}s_{i+2m}+2\lambda_{b}s_{i}-\lambda_{b}\right\}\\
		&=0,
	\end{aligned}
\end{equation}

\begin{equation}
	\begin{aligned}
		&\mathrm{E}\left\{\operatorname{Poi}\left(\left(\frac{k}{n}-\varepsilon\right)\left(\lambda_{s} s_{i+2m+1}+\lambda_{b}\right)\right) \times\left(2 s_{i}-1\right)\right\} \\
		&=\mathrm{E}\left\{\mathrm{E}\left[\operatorname{Poi}\left(\left(\frac{k}{n}-\varepsilon\right)\left(\lambda_{s} s_{i+2m+1}+\lambda_{b}\right)\right) \times\left(2 s_{i}-1\right)\mid \bm{s}\right]\right\} \\
		&=\left(\frac{k}{n}-\varepsilon\right)\mathrm{E}\left\{2\lambda_{s}s_{i}s_{i+2m+1}-\lambda_{s}s_{i+2m+1}+2\lambda_{b}s_{i}-\lambda_{b}\right\}\\
		&=0.
	\end{aligned}
\end{equation}

Thus, 
\begin{equation}
	\mathrm{E}\left\{n_{i} \times\left(2 s_{i}-1\right)\right\}=0, i=L-2m.
\end{equation}

We have $ s_{i+2m} = 0$ as well as $ s_{i+2m+1} = 0$ when $(L-2m+1) \leq i \leq L $. There are $ 2m $ cumulative item in the $ \operatorname{C}_{2m,k} $ follows
\begin{equation}
	\begin{aligned}
		&\mathrm{E}\left\{\operatorname{Poi}\left(\left(1-\frac{k}{n}\right)\left(\lambda_{s} s_{i+2m}+\lambda_{b}\right)\right) \times\left(2 s_{i}-1\right)\right\} \\
		&=\mathrm{E}\left\{\mathrm{E}\left[\operatorname{Poi}\left(\left(1-\frac{k}{n}\right)\left(\lambda_{s} s_{i+2m}+\lambda_{b}\right)\right) \times\left(2 s_{i}-1\right)\mid \bm{s}\right]\right\} \\
		&=(1-\frac{k}{n})\mathrm{E}\left\{2\lambda_{s}s_{i}s_{i+2m}-\lambda_{s}s_{i+2m}+2\lambda_{b}s_{i}-\lambda_{b}\right\}\\
		&=0,
	\end{aligned}
\end{equation}

\begin{equation}
	\begin{aligned}
		&\mathrm{E}\left\{\operatorname{Poi}\left(\varepsilon\left(\lambda_{s} s_{i+2m}+\lambda_{b}\right)\right) \times\left(2 s_{i}-1\right)\right\}  \\
		&=\mathrm{E}\left\{\mathrm{E}\left[\operatorname{Poi}\left(\varepsilon\left(\lambda_{s} s_{i+2m}+\lambda_{b}\right)\right) \times\left(2 s_{i}-1\right)\mid \bm{s}\right]\right\} \\
		&=\varepsilon\mathrm{E}\left\{2\lambda_{s}s_{i}s_{i+2m}-\lambda_{s}s_{i+2m}+2\lambda_{b}s_{i}-\lambda_{b}\right\}\\
		&=0,
	\end{aligned}
\end{equation}

\begin{equation}
	\begin{aligned}
		&\mathrm{E}\left\{\operatorname{Poi}\left(\left(\frac{k}{n}-\varepsilon\right)\left(\lambda_{s} s_{i+2m+1}+\lambda_{b}\right)\right) \times\left(2 s_{i}-1\right)\right\} \\
		&=\mathrm{E}\left\{\mathrm{E}\left[\operatorname{Poi}\left(\left(\frac{k}{n}-\varepsilon\right)\left(\lambda_{s} s_{i+2m+1}+\lambda_{b}\right)\right) \times\left(2 s_{i}-1\right)\mid \bm{s}\right]\right\} \\
		&=\left(\frac{k}{n}-\varepsilon\right)\mathrm{E}\left\{2\lambda_{s}s_{i}s_{i+2m+1}-\lambda_{s}s_{i+2m+1}+2\lambda_{b}s_{i}-\lambda_{b}\right\}\\
		&=0.
	\end{aligned}
\end{equation}

Thus, 
\begin{equation}
	\mathrm{E}\left\{n_{i} \times\left(2 s_{i}-1\right)\right\}=0, (L-2m+1) \leq i \leq L.
\end{equation}

If $ s_{i} $ lies in RANDOM part, there are $ \left(1-\alpha\right)L $ cumulative items in the $ \operatorname{C}_{2m,k} $ in total follows
\begin{equation}
	\mathrm{E}\left\{n_{i} \times\left(2 s_{i}-1\right)\right\}=0.
\end{equation}

On the other hand, $ s_i$ lies in $\{1,0,...\} $ part. We have $ \mathrm{E}\left(s_{i}s_{i+2m}\right) = 1$ and $ \mathrm{E}\left(s_{i}s_{i+2m+1}\right) = 0$ when $ \left(\frac{1-\alpha}{2} L +2\right) \leq i \leq \left(\frac{1+\alpha}{2}L-2m-1\right) $. There are $ \alpha L-2m-2 $ cumulative items in the $ \operatorname{C}_{2m,k} $ follows

\begin{equation}
	\begin{aligned}
		&\mathrm{E}\left\{\operatorname{Poi}\left(\left(1-\frac{k}{n}\right)\left(\lambda_{s} s_{i+2m}+\lambda_{b}\right)\right) \times\left(2 s_{i}-1\right)\right\} \\
		&=\mathrm{E}\left\{\mathrm{E}\left[\operatorname{Poi}\left(\left(1-\frac{k}{n}\right)\left(\lambda_{s} s_{i+2m}+\lambda_{b}\right)\right) \times\left(2 s_{i}-1\right)\mid \bm{s}\right]\right\} \\
		&=(1-\frac{k}{n})\mathrm{E}\left\{2\lambda_{s}s_{i}s_{i+2m}-\lambda_{s}s_{i+2m}+2\lambda_{b}s_{i}-\lambda_{b}\right\}\\
		&=(1-\frac{k}{n}) \frac{\lambda_{s}}{2},
	\end{aligned}
\end{equation}

\begin{equation}
	\begin{aligned}
		&\mathrm{E}\left\{\operatorname{Poi}\left(\varepsilon\left(\lambda_{s} s_{i+2m}+\lambda_{b}\right)\right) \times\left(2 s_{i}-1\right)\right\}  \\
		&=\mathrm{E}\left\{\mathrm{E}\left[\operatorname{Poi}\left(\varepsilon\left(\lambda_{s} s_{i+2m}+\lambda_{b}\right)\right) \times\left(2 s_{i}-1\right)\mid \bm{s}\right]\right\} \\
		&=\varepsilon\mathrm{E}\left\{2\lambda_{s}s_{i}s_{i+2m}-\lambda_{s}s_{i+2m}+2\lambda_{b}s_{i}-\lambda_{b}\right\}\\
		&=\varepsilon \frac{\lambda_{s}}{2},
	\end{aligned}
\end{equation}

\begin{equation}
	\begin{aligned}
		&\mathrm{E}\left\{\operatorname{Poi}\left(\left(\frac{k}{n}-\varepsilon\right)\left(\lambda_{s} s_{i+2m+1}+\lambda_{b}\right)\right) \times\left(2 s_{i}-1\right)\right\} \\
		&=\mathrm{E}\left\{\mathrm{E}\left[\operatorname{Poi}\left(\left(\frac{k}{n}-\varepsilon\right)\left(\lambda_{s} s_{i+2m+1}+\lambda_{b}\right)\right) \times\left(2 s_{i}-1\right)\mid \bm{s}\right]\right\} \\
		&=\left(\frac{k}{n}-\varepsilon\right)\mathrm{E}\left\{2\lambda_{s}s_{i}s_{i+2m+1}-\lambda_{s}s_{i+2m+1}+2\lambda_{b}s_{i}-\lambda_{b}\right\}\\
		&=-(\frac{k}{n}-\varepsilon) \frac{\lambda_{s}}{2}.
	\end{aligned}
\end{equation}

If $ s_i$ lies in $\{1,0,...\} $ part, we have $ s_{i}=s_{i+2m}=0$ and $ \mathrm{E}\left(s_{i}s_{i+2m+1}\right) = 0$ when $  i = \left(\frac{1+\alpha}{2} L-2m\right) $. $ s_{i+2m} $ is the end of $\{1,0,...\} $ part. There are $ 1 $ cumulative item in the $ \operatorname{C}_{2m,k} $ follows
\begin{equation}
	\begin{aligned}
		&\mathrm{E}\left\{\operatorname{Poi}\left(\left(1-\frac{k}{n}\right)\left(\lambda_{s} s_{i+2m}+\lambda_{b}\right)\right) \times\left(2 s_{i}-1\right)\right\} \\
		&=\mathrm{E}\left\{\mathrm{E}\left[\operatorname{Poi}\left(\left(1-\frac{k}{n}\right)\left(\lambda_{s} s_{i+2m}+\lambda_{b}\right)\right) \times\left(2 s_{i}-1\right)\mid \bm{s}\right]\right\} \\
		&=(1-\frac{k}{n})\mathrm{E}\left\{2\lambda_{s}s_{i}s_{i+2m}-\lambda_{s}s_{i+2m}+2\lambda_{b}s_{i}-\lambda_{b}\right\}\\
		&=-(1-\frac{k}{n}) \lambda_{b},
	\end{aligned}
\end{equation}

\begin{equation}
	\begin{aligned}
		&\mathrm{E}\left\{\operatorname{Poi}\left(\varepsilon\left(\lambda_{s} s_{i+2m}+\lambda_{b}\right)\right) \times\left(2 s_{i}-1\right)\right\}  \\
		&=\mathrm{E}\left\{\mathrm{E}\left[\operatorname{Poi}\left(\varepsilon\left(\lambda_{s} s_{i+2m}+\lambda_{b}\right)\right) \times\left(2 s_{i}-1\right)\mid \bm{s}\right]\right\} \\
		&=\varepsilon\mathrm{E}\left\{2\lambda_{s}s_{i}s_{i+2m}-\lambda_{s}s_{i+2m}+2\lambda_{b}s_{i}-\lambda_{b}\right\}\\
		&=-\varepsilon \lambda_{b},
	\end{aligned}
\end{equation}

\begin{equation}
	\begin{aligned}
		&\mathrm{E}\left\{\operatorname{Poi}\left(\left(\frac{k}{n}-\varepsilon\right)\left(\lambda_{s} s_{i+2m+1}+\lambda_{b}\right)\right) \times\left(2 s_{i}-1\right)\right\} \\
		&=\mathrm{E}\left\{\mathrm{E}\left[\operatorname{Poi}\left(\left(\frac{k}{n}-\varepsilon\right)\left(\lambda_{s} s_{i+2m+1}+\lambda_{b}\right)\right) \times\left(2 s_{i}-1\right)\mid \bm{s}\right]\right\} \\
		&=\left(\frac{k}{n}-\varepsilon\right)\mathrm{E}\left\{2\lambda_{s}s_{i}s_{i+2m+1}-\lambda_{s}s_{i+2m+1}+2\lambda_{b}s_{i}-\lambda_{b}\right\}\\
		&=-\left(\frac{k}{n}-\varepsilon\right)(\frac{\lambda_{s}}{2}+\lambda_{b}).
	\end{aligned}
\end{equation}

Paired with the boundary above, we have $ s_{i}=s_{i+2m}=1$ and $ \mathrm{E}\left(s_{i}s_{i+2m+1}\right) = 0$ when $  i = \left(\frac{1-\alpha}{2} L +1\right) $. $ s_{i} $ is the start of $\{1,0,...\} $ part. There are $ 1 $ cumulative item in the $ \operatorname{C}_{2m,k} $ follows

\begin{equation}
	\begin{aligned}
		&\mathrm{E}\left\{\operatorname{Poi}\left(\left(1-\frac{k}{n}\right)\left(\lambda_{s} s_{i+2m}+\lambda_{b}\right)\right) \times\left(2 s_{i}-1\right)\right\} \\
		&=\mathrm{E}\left\{\mathrm{E}\left[\operatorname{Poi}\left(\left(1-\frac{k}{n}\right)\left(\lambda_{s} s_{i+2m}+\lambda_{b}\right)\right) \times\left(2 s_{i}-1\right)\mid \bm{s}\right]\right\} \\
		&=(1-\frac{k}{n})\mathrm{E}\left\{2\lambda_{s}s_{i}s_{i+2m}-\lambda_{s}s_{i+2m}+2\lambda_{b}s_{i}-\lambda_{b}\right\}\\
		&=(1-\frac{k}{n}) (\lambda_{b}+\lambda_{b}),
	\end{aligned}
\end{equation}

\begin{equation}
	\begin{aligned}
		&\mathrm{E}\left\{\operatorname{Poi}\left(\varepsilon\left(\lambda_{s} s_{i+2m}+\lambda_{b}\right)\right) \times\left(2 s_{i}-1\right)\right\}  \\
		&=\mathrm{E}\left\{\mathrm{E}\left[\operatorname{Poi}\left(\varepsilon\left(\lambda_{s} s_{i+2m}+\lambda_{b}\right)\right) \times\left(2 s_{i}-1\right)\mid \bm{s}\right]\right\} \\
		&=\varepsilon\mathrm{E}\left\{2\lambda_{s}s_{i}s_{i+2m}-\lambda_{s}s_{i+2m}+2\lambda_{b}s_{i}-\lambda_{b}\right\}\\
		&=\varepsilon (\lambda_{s}+\lambda_{b}),
	\end{aligned}
\end{equation}

\begin{equation}
	\begin{aligned}
		&\mathrm{E}\left\{\operatorname{Poi}\left(\left(\frac{k}{n}-\varepsilon\right)\left(\lambda_{s} s_{i+2m+1}+\lambda_{b}\right)\right) \times\left(2 s_{i}-1\right)\right\} \\
		&=\mathrm{E}\left\{\mathrm{E}\left[\operatorname{Poi}\left(\left(\frac{k}{n}-\varepsilon\right)\left(\lambda_{s} s_{i+2m+1}+\lambda_{b}\right)\right) \times\left(2 s_{i}-1\right)\mid \bm{s}\right]\right\} \\
		&=\left(\frac{k}{n}-\varepsilon\right)\mathrm{E}\left\{2\lambda_{s}s_{i}s_{i+2m+1}-\lambda_{s}s_{i+2m+1}+2\lambda_{b}s_{i}-\lambda_{b}\right\}\\
		&=\left(\frac{k}{n}-\varepsilon\right)\lambda_{b}.
	\end{aligned}
\end{equation}

If $ s_i$ lies in $\{1,0,...\} $ part, we have $ \mathrm{E}\left(s_{i}s_{i+2m}\right) = \frac{1}{4}$ and $ \mathrm{E}\left(s_{i}s_{i+2m+1}\right) = \frac{1}{4}$ when $ \left(\frac{1+\alpha}{2} L-2m+1 \leq i \leq \frac{1+\alpha}{2} L\right) $. There are $ 2m $ cumulative items in the $ \operatorname{C}_{2m,k} $ follows

\begin{equation}
	\mathrm{E}\left\{n_{i} \times\left(2 s_{i}-1\right)\right\}=0.
\end{equation}

Thus, the expectation of  $ \operatorname{C}_{2m,k} $ is,
\begin{align}
	\mathrm{E}\left(\operatorname{C}_{2m, k}\right)&=(\alpha L-2 m-2) \frac{\lambda_{s}}{2}\left[1-2\left(\frac{k}{n}-\varepsilon\right)\right]-\lambda_{b}+\lambda_{b}+  \frac{\lambda_{s}}{2}\left[1-3\left(\frac{k}{n}-\varepsilon\right)\right]\\
	&=\frac{\lambda_{s}}{2}\left\{(\alpha L-2 m)\left[1-2\left(\frac{k}{n}-\varepsilon\right)\right]+\left(\frac{k}{n}-\varepsilon\right)\right\},~~2m \leq \min\{\alpha L, \frac{1-\alpha}{2}L\}.
\end{align}

Similar to  of Case 1, the expectation of variance of $ \operatorname{C}_{2m,k}  $ is,
\begin{align}
	\mathrm{E}\left(\operatorname{Var}\left(\operatorname{C}_{2m,k}\right)\right)=\left(L-2m\right)\left(\frac{\lambda_{s}}{2}+\lambda_{b}\right),~~2m \leq \min\{\alpha L, \frac{1-\alpha}{2}L\}.
\end{align}

Compared to Case 1, the cumulative terms from  missed $ 2m $ symbols in the variance calculation are ruled out.

\section{ Proof of Theorem \ref{theorem.covforCase3} }  \label{appendix.covforCase3}
Similar to Appendix \ref{appendix.covforCase2}, the covariance is mainly composed of covariance of Poisson random variables, such as $ \operatorname{Cov}\left\{\left(2 s_{i}-1\right) \operatorname{Poi}\left(\varepsilon\left(\lambda_{s} s_{i-1}+\lambda_{b}\right)\right),\left(2 s_{j}-1\right) \operatorname{Poi}\left(\varepsilon\left(\lambda_{s} s_{j+2m}+\lambda_{b}\right)\right)\right\} $ when $ i-1=j+2m $, which means $\operatorname{Poi}\left(\varepsilon\left(\lambda_{s} s_{i-1}+\lambda_{b}\right)\right)$ and $ \operatorname{Poi}\left(\varepsilon\left(\lambda_{s} s_{j+2m}+\lambda_{b}\right)\right) $ are actually the same random variable as a result that they are from Poisson process during the same period of time.

First, $ s_j$ lies in RANDOM part. We have $ \mathrm{E}\left(s_{j}s_{j+2m}\right) = \frac{1}{4}$ and $ \mathrm{E}\left(s_{j}s_{j+2m+1}\right) = \frac{1}{4}$ and $\mathrm{E}$$\left\{s_{j}s_{j+2m}s_{j+2m+1}\right\}=\frac{1}{8}$ when $1 \leq j \leq (\frac{1-\alpha}{2} L-2m-1) $ or $ (\frac{1+\alpha}{2} L+1)\leq j \leq \left(L-2m-1\right) $. $ s_{j},s_{j+2m},s_{j+2m+1} $ lie in RANDOM part.


For $ \left((1-\alpha)-4m-2\right) $ pairs of $ \left(i,j\right) $ with $ i-1=j+2m $, 
\begin{equation}
	\begin{aligned}
	&\mathrm{E}\left(\operatorname{Cov}\left\{\left(2 s_{i}-1\right) \operatorname{Poi}\left(\varepsilon\left(\lambda_{s} s_{i-1}+\lambda_{b}\right)\right),\left(2 s_{j}-1\right) \operatorname{Poi}\left(\varepsilon\left(\lambda_{s} s_{j+2m}+\lambda_{b}\right)\right)\right\}\right)\\
	&=\mathrm{E}\left\{\mathrm{E}\left(\operatorname{Cov}\left\{\left(2 s_{i}-1\right) \operatorname{Poi}\left(\varepsilon\left(\lambda_{s} s_{i-1}+\lambda_{b}\right)\right),\left(2 s_{j}-1\right) \operatorname{Poi}\left(\varepsilon\left(\lambda_{s} s_{j+2m}+\lambda_{b}\right)\right)\right\}\right)\mid \bm{s} \right\}  \\
	&=\mathrm{E}\left\{\left(2 s_{j}-1\right)\left(2 s_{j+2m+1}-1\right)\left(\varepsilon\left(\lambda_{s} s_{j+2m}+\lambda_{b}\right)\right)\right\}\\
	&=0,\\
	\end{aligned}
\end{equation}

\begin{equation}
	\begin{aligned}
		&\mathrm{E}\left(\operatorname{Cov}\left\{\left(2 s_{i}-1\right) \operatorname{Poi}\left(\left(\frac{k}{n}-\varepsilon\right)\left(\lambda_{s} s_{i}+\lambda_{b}\right)\right),\left(2 s_{j}-1\right) \operatorname{Poi}\left(\left(\frac{k}{n}-\varepsilon\right)\left(\lambda_{s} s_{j+2m+1}+\lambda_{b}\right)\right)\right\}\right)\\
		&=\mathrm{E}\left\{\mathrm{E}\left(\operatorname{Cov}\left\{\left(2 s_{i}-1\right) \operatorname{Poi}\left(\left(\frac{k}{n}-\varepsilon\right)\left(\lambda_{s} s_{i}+\lambda_{b}\right)\right),\left(2 s_{j}-1\right) \operatorname{Poi}\left(\left(\frac{k}{n}-\varepsilon\right)\left(\lambda_{s} s_{j+2m+1}+\lambda_{b}\right)\right)\right\}\right)\mid \bm{s} \right\}  \\
		&=\mathrm{E}\left\{\left(2 s_{j}-1\right)\left(2 s_{j+2m+1}-1\right)\left(\left(\frac{k}{n}-\varepsilon\right)\left(\lambda_{s} s_{j+2m+1}+\lambda_{b}\right)\right)\right\}\\
		&=0,
	\end{aligned}
\end{equation}


For $ \left((1-\alpha)-4m-2\right) $ pairs of $ \left(i,j\right) $ with $ i=j+2m $,
\begin{equation}
	\begin{aligned}
		&\mathrm{E}\left(\operatorname{Cov}\left\{\left(2 s_{i}-1\right) \operatorname{Poi}\left(\left(1-\frac{k}{n}\right)\left(\lambda_{s} s_{i}+\lambda_{b}\right)\right),\left(2 s_{j}-1\right) \operatorname{Poi}\left(\left(1-\frac{k}{n}\right)\left(\lambda_{s} s_{j+2m}+\lambda_{b}\right)\right)\right\}\right)\\
		&=\mathrm{E}\left\{\mathrm{E}\left(\operatorname{Cov}\left\{\left(2 s_{i}-1\right) \operatorname{Poi}\left(\left(1-\frac{k}{n}\right)\left(\lambda_{s} s_{i}+\lambda_{b}\right)\right),\left(2 s_{j}-1\right) \operatorname{Poi}\left(\left(1-\frac{k}{n}\right)\left(\lambda_{s} s_{j+2m}+\lambda_{b}\right)\right)\right\}\right)\mid \bm{s} \right\}  \\
		&=\mathrm{E}\left\{\left(2 s_{j}-1\right)\left(2 s_{j+2m}-1\right)\left(\left(1-\frac{k}{n}\right)\left(\lambda_{s} s_{j+2m}+\lambda_{b}\right)\right)\right\}\\
		&=0,
	\end{aligned}
\end{equation}

We have $ s_{j+2m} $ lying in RANDOM part as well as $ s_{j+2m+1} = 1$ when $ j=\frac{1-\alpha}{2}L-2m $. $ s_{j+2m+1} $ lies in $\{1,0,...\} $ part.

For $ 1 $ pair of $ \left(i,j\right) $ with $ i-1=j+2m $, 
\begin{equation}
	\begin{aligned}
		&\mathrm{E}\left(\operatorname{Cov}\left\{\left(2 s_{i}-1\right) \operatorname{Poi}\left(\varepsilon\left(\lambda_{s} s_{i-1}+\lambda_{b}\right)\right),\left(2 s_{j}-1\right) \operatorname{Poi}\left(\varepsilon\left(\lambda_{s} s_{j+2m}+\lambda_{b}\right)\right)\right\}\right)\\
		&=\mathrm{E}\left\{\mathrm{E}\left(\operatorname{Cov}\left\{\left(2 s_{i}-1\right) \operatorname{Poi}\left(\varepsilon\left(\lambda_{s} s_{i-1}+\lambda_{b}\right)\right),\left(2 s_{j}-1\right) \operatorname{Poi}\left(\varepsilon\left(\lambda_{s} s_{j+2m}+\lambda_{b}\right)\right)\right\}\right)\mid \bm{s} \right\}  \\
		&=\mathrm{E}\left\{\left(2 s_{j}-1\right)\left(2 s_{j+2m+1}-1\right)\left(\varepsilon\left(\lambda_{s} s_{j+2m}+\lambda_{b}\right)\right)\right\}\\
		&=\varepsilon\mathrm{E}\left(4\lambda_{s}s_{j}s_{j+2m}s_{j+2m+1}-2\lambda_{s}s_{j}s_{j+2m}-2\lambda_{s}s_{j+2m}s_{j+2m+1}+\lambda_{s}s_{j+2m}\right)\\
		&+\varepsilon\mathrm{E}\left(4\lambda_{b}s_{j}s_{j+2m+1}-2\lambda_{b}s_{j}-2\lambda_{b}s_{j+2m+1}+\lambda_{b}\right)\\
		&=0,\\
	\end{aligned}
\end{equation}

\begin{equation}
	\begin{aligned}
		&\mathrm{E}\left(\operatorname{Cov}\left\{\left(2 s_{i}-1\right) \operatorname{Poi}\left(\left(\frac{k}{n}-\varepsilon\right)\left(\lambda_{s} s_{i}+\lambda_{b}\right)\right),\left(2 s_{j}-1\right) \operatorname{Poi}\left(\left(\frac{k}{n}-\varepsilon\right)\left(\lambda_{s} s_{j+2m+1}+\lambda_{b}\right)\right)\right\}\right)\\
		&=\mathrm{E}\left\{\mathrm{E}\left(\operatorname{Cov}\left\{\left(2 s_{i}-1\right) \operatorname{Poi}\left(\left(\frac{k}{n}-\varepsilon\right)\left(\lambda_{s} s_{i}+\lambda_{b}\right)\right),\left(2 s_{j}-1\right) \operatorname{Poi}\left(\left(\frac{k}{n}-\varepsilon\right)\left(\lambda_{s} s_{j+2m+1}+\lambda_{b}\right)\right)\right\}\right)\mid \bm{s} \right\}  \\
		&=\mathrm{E}\left\{\left(2 s_{j}-1\right)\left(2 s_{j+2m+1}-1\right)\left(\left(\frac{k}{n}-\varepsilon\right)\left(\lambda_{s} s_{j+2m+1}+\lambda_{b}\right)\right)\right\}\\
		&=\left(\frac{k}{n}-\varepsilon\right)\mathrm{E}\left(4\lambda_{s}s_{j}s_{j+2m+1}^2-2\lambda_{s}s_{j}s_{j+2m+1}-2\lambda_{s}s_{j+2m+1}^2+\lambda_{s}s_{j+2m+1}\right)\\
		&+\left(\frac{k}{n}-\varepsilon\right)\mathrm{E}\left(4\lambda_{b}s_{j}s_{j+2m+1}-2\lambda_{b}s_{j}-2\lambda_{b}s_{j+2m+1}+\lambda_{b}\right)\\
		&=0,
	\end{aligned}
\end{equation}

For $ 1 $ pair of $ \left(i,j\right) $ with $ i=j+2m $,
\begin{equation}
	\begin{aligned}
		&\mathrm{E}\left(\operatorname{Cov}\left\{\left(2 s_{i}-1\right) \operatorname{Poi}\left(\left(1-\frac{k}{n}\right)\left(\lambda_{s} s_{i}+\lambda_{b}\right)\right),\left(2 s_{j}-1\right) \operatorname{Poi}\left(\left(1-\frac{k}{n}\right)\left(\lambda_{s} s_{j+2m}+\lambda_{b}\right)\right)\right\}\right)\\
		&=\mathrm{E}\left\{\mathrm{E}\left(\operatorname{Cov}\left\{\left(2 s_{i}-1\right) \operatorname{Poi}\left(\left(1-\frac{k}{n}\right)\left(\lambda_{s} s_{i}+\lambda_{b}\right)\right),\left(2 s_{j}-1\right) \operatorname{Poi}\left(\left(1-\frac{k}{n}\right)\left(\lambda_{s} s_{j+2m}+\lambda_{b}\right)\right)\right\}\right)\mid \bm{s} \right\}  \\
		&=\mathrm{E}\left\{\left(2 s_{j}-1\right)\left(2 s_{j+2m}-1\right)\left(\left(1-\frac{k}{n}\right)\left(\lambda_{s} s_{j+2m}+\lambda_{b}\right)\right)\right\}\\
		&=\left(1-\frac{k}{n}\right)\mathrm{E}\left(4\lambda_{s}s_{j}s_{j+2m}^2-2\lambda_{s}s_{j}s_{j+2m}-2\lambda_{s}s_{j+2m}^2+\lambda_{s}s_{j+2m}\right)\\
		&+\left(1-\frac{k}{n}\right)\mathrm{E}\left(4\lambda_{b}s_{j}s_{j+2m}-2\lambda_{b}s_{j}-2\lambda_{b}s_{j+2m}+\lambda_{b}\right)\\
		&=0,
	\end{aligned}
\end{equation}

We have $ s_{j+2m} $ and $ s_{j+2m+1} $ lying in $\{1,0,...\} $ part when $ s_{j} $ lies in RANDOM part and $(\frac{1-\alpha}{2}L-2m+1) \leq j \leq \frac{1-\alpha}{2}L$. Thus, $  s_{j+2m} s_{j+2m+1} =0 $.

For $ 2m $ pair of $ \left(i,j\right) $ with $ i-1=j+2m $, 
\begin{equation}
	\begin{aligned}
		&\mathrm{E}\left(\operatorname{Cov}\left\{\left(2 s_{i}-1\right) \operatorname{Poi}\left(\varepsilon\left(\lambda_{s} s_{i-1}+\lambda_{b}\right)\right),\left(2 s_{j}-1\right) \operatorname{Poi}\left(\varepsilon\left(\lambda_{s} s_{j+2m}+\lambda_{b}\right)\right)\right\}\right)\\
		&=\mathrm{E}\left\{\mathrm{E}\left(\operatorname{Cov}\left\{\left(2 s_{i}-1\right) \operatorname{Poi}\left(\varepsilon\left(\lambda_{s} s_{i-1}+\lambda_{b}\right)\right),\left(2 s_{j}-1\right) \operatorname{Poi}\left(\varepsilon\left(\lambda_{s} s_{j+2m}+\lambda_{b}\right)\right)\right\}\right)\mid \bm{s} \right\}  \\
		&=\mathrm{E}\left\{\left(2 s_{j}-1\right)\left(2 s_{j+2m+1}-1\right)\left(\varepsilon\left(\lambda_{s} s_{j+2m}+\lambda_{b}\right)\right)\right\}\\
		&=\varepsilon\mathrm{E}\left(4\lambda_{s}s_{j}s_{j+2m}s_{j+2m+1}-2\lambda_{s}s_{j}s_{j+2m}-2\lambda_{s}s_{j+2m}s_{j+2m+1}+\lambda_{s}s_{j+2m}\right)\\
		&+\varepsilon\mathrm{E}\left(4\lambda_{b}s_{j}s_{j+2m+1}-2\lambda_{b}s_{j}-2\lambda_{b}s_{j+2m+1}+\lambda_{b}\right)\\
		&=0,\\
	\end{aligned}
\end{equation}

\begin{equation}
	\begin{aligned}
		&\mathrm{E}\left(\operatorname{Cov}\left\{\left(2 s_{i}-1\right) \operatorname{Poi}\left(\left(\frac{k}{n}-\varepsilon\right)\left(\lambda_{s} s_{i}+\lambda_{b}\right)\right),\left(2 s_{j}-1\right) \operatorname{Poi}\left(\left(\frac{k}{n}-\varepsilon\right)\left(\lambda_{s} s_{j+2m+1}+\lambda_{b}\right)\right)\right\}\right)\\
		&=\mathrm{E}\left\{\mathrm{E}\left(\operatorname{Cov}\left\{\left(2 s_{i}-1\right) \operatorname{Poi}\left(\left(\frac{k}{n}-\varepsilon\right)\left(\lambda_{s} s_{i}+\lambda_{b}\right)\right),\left(2 s_{j}-1\right) \operatorname{Poi}\left(\left(\frac{k}{n}-\varepsilon\right)\left(\lambda_{s} s_{j+2m+1}+\lambda_{b}\right)\right)\right\}\right)\mid \bm{s} \right\}  \\
		&=\mathrm{E}\left\{\left(2 s_{j}-1\right)\left(2 s_{j+2m+1}-1\right)\left(\left(\frac{k}{n}-\varepsilon\right)\left(\lambda_{s} s_{j+2m+1}+\lambda_{b}\right)\right)\right\}\\
		&=\left(\frac{k}{n}-\varepsilon\right)\mathrm{E}\left(4\lambda_{s}s_{j}s_{j+2m+1}^2-2\lambda_{s}s_{j}s_{j+2m+1}-2\lambda_{s}s_{j+2m+1}^2+\lambda_{s}s_{j+2m+1}\right)\\
		&+\left(\frac{k}{n}-\varepsilon\right)\mathrm{E}\left(4\lambda_{b}s_{j}s_{j+2m+1}-2\lambda_{b}s_{j}-2\lambda_{b}s_{j+2m+1}+\lambda_{b}\right)\\
		&=0,
	\end{aligned}
\end{equation}

For $ 2m $ pair of $ \left(i,j\right) $ with $ i=j+2m $,
\begin{equation}
	\begin{aligned}
		&\mathrm{E}\left(\operatorname{Cov}\left\{\left(2 s_{i}-1\right) \operatorname{Poi}\left(\left(1-\frac{k}{n}\right)\left(\lambda_{s} s_{i}+\lambda_{b}\right)\right),\left(2 s_{j}-1\right) \operatorname{Poi}\left(\left(1-\frac{k}{n}\right)\left(\lambda_{s} s_{j+2m}+\lambda_{b}\right)\right)\right\}\right)\\
		&=\mathrm{E}\left\{\mathrm{E}\left(\operatorname{Cov}\left\{\left(2 s_{i}-1\right) \operatorname{Poi}\left(\left(1-\frac{k}{n}\right)\left(\lambda_{s} s_{i}+\lambda_{b}\right)\right),\left(2 s_{j}-1\right) \operatorname{Poi}\left(\left(1-\frac{k}{n}\right)\left(\lambda_{s} s_{j+2m}+\lambda_{b}\right)\right)\right\}\right)\mid \bm{s} \right\}  \\
		&=\mathrm{E}\left\{\left(2 s_{j}-1\right)\left(2 s_{j+2m}-1\right)\left(\left(1-\frac{k}{n}\right)\left(\lambda_{s} s_{j+2m}+\lambda_{b}\right)\right)\right\}\\
		&=\left(1-\frac{k}{n}\right)\mathrm{E}\left(4\lambda_{s}s_{j}s_{j+2m}^2-2\lambda_{s}s_{j}s_{j+2m}-2\lambda_{s}s_{j+2m}^2+\lambda_{s}s_{j+2m}\right)\\
		&+\left(1-\frac{k}{n}\right)\mathrm{E}\left(4\lambda_{b}s_{j}s_{j+2m}-2\lambda_{b}s_{j}-2\lambda_{b}s_{j+2m}+\lambda_{b}\right)\\
		&=0,
	\end{aligned}
\end{equation}

We have $ s_{j} $ and $ s_{j+2m} $ lying in RANDOM part as well as $ s_{j+2m+1} = 0$ when $ j=L-2m $.

For $ 1 $ pair of $ \left(i,j\right) $ with $ i-1=j+2m $, 
\begin{equation}
	\begin{aligned}
		&\mathrm{E}\left(\operatorname{Cov}\left\{\left(2 s_{i}-1\right) \operatorname{Poi}\left(\varepsilon\left(\lambda_{s} s_{i-1}+\lambda_{b}\right)\right),\left(2 s_{j}-1\right) \operatorname{Poi}\left(\varepsilon\left(\lambda_{s} s_{j+2m}+\lambda_{b}\right)\right)\right\}\right)\\
		&=\mathrm{E}\left\{\mathrm{E}\left(\operatorname{Cov}\left\{\left(2 s_{i}-1\right) \operatorname{Poi}\left(\varepsilon\left(\lambda_{s} s_{i-1}+\lambda_{b}\right)\right),\left(2 s_{j}-1\right) \operatorname{Poi}\left(\varepsilon\left(\lambda_{s} s_{j+2m}+\lambda_{b}\right)\right)\right\}\right)\mid \bm{s} \right\}  \\
		&=\mathrm{E}\left\{\left(2 s_{j}-1\right)\left(2 s_{j+2m+1}-1\right)\left(\varepsilon\left(\lambda_{s} s_{j+2m}+\lambda_{b}\right)\right)\right\}\\
		&=\varepsilon\mathrm{E}\left(4\lambda_{s}s_{j}s_{j+2m}s_{j+2m+1}-2\lambda_{s}s_{j}s_{j+2m}-2\lambda_{s}s_{j+2m}s_{j+2m+1}+\lambda_{s}s_{j+2m}\right)\\
		&+\varepsilon\mathrm{E}\left(4\lambda_{b}s_{j}s_{j+2m+1}-2\lambda_{b}s_{j}-2\lambda_{b}s_{j+2m+1}+\lambda_{b}\right)\\
		&=0,\\
	\end{aligned}
\end{equation}

\begin{equation}
	\begin{aligned}
		&\mathrm{E}\left(\operatorname{Cov}\left\{\left(2 s_{i}-1\right) \operatorname{Poi}\left(\left(\frac{k}{n}-\varepsilon\right)\left(\lambda_{s} s_{i}+\lambda_{b}\right)\right),\left(2 s_{j}-1\right) \operatorname{Poi}\left(\left(\frac{k}{n}-\varepsilon\right)\left(\lambda_{s} s_{j+2m+1}+\lambda_{b}\right)\right)\right\}\right)\\
		&=\mathrm{E}\left\{\mathrm{E}\left(\operatorname{Cov}\left\{\left(2 s_{i}-1\right) \operatorname{Poi}\left(\left(\frac{k}{n}-\varepsilon\right)\left(\lambda_{s} s_{i}+\lambda_{b}\right)\right),\left(2 s_{j}-1\right) \operatorname{Poi}\left(\left(\frac{k}{n}-\varepsilon\right)\left(\lambda_{s} s_{j+2m+1}+\lambda_{b}\right)\right)\right\}\right)\mid \bm{s} \right\}  \\
		&=\mathrm{E}\left\{\left(2 s_{j}-1\right)\left(2 s_{j+2m+1}-1\right)\left(\left(\frac{k}{n}-\varepsilon\right)\left(\lambda_{s} s_{j+2m+1}+\lambda_{b}\right)\right)\right\}\\
		&=\left(\frac{k}{n}-\varepsilon\right)\mathrm{E}\left(4\lambda_{s}s_{j}s_{j+2m+1}^2-2\lambda_{s}s_{j}s_{j+2m+1}-2\lambda_{s}s_{j+2m+1}^2+\lambda_{s}s_{j+2m+1}\right)\\
		&+\left(\frac{k}{n}-\varepsilon\right)\mathrm{E}\left(4\lambda_{b}s_{j}s_{j+2m+1}-2\lambda_{b}s_{j}
		-2\lambda_{b}s_{j+2m+1}+\lambda_{b}\right)\\
		&=0,
	\end{aligned}
\end{equation}

For $ 1 $ pair of $ \left(i,j\right) $ with $ i=j+2m $,
\begin{equation}
	\begin{aligned}
		&\mathrm{E}\left(\operatorname{Cov}\left\{\left(2 s_{i}-1\right) \operatorname{Poi}\left(\left(1-\frac{k}{n}\right)\left(\lambda_{s} s_{i}+\lambda_{b}\right)\right),\left(2 s_{j}-1\right) \operatorname{Poi}\left(\left(1-\frac{k}{n}\right)\left(\lambda_{s} s_{j+2m}+\lambda_{b}\right)\right)\right\}\right)\\
		&=\mathrm{E}\left\{\mathrm{E}\left(\operatorname{Cov}\left\{\left(2 s_{i}-1\right) \operatorname{Poi}\left(\left(1-\frac{k}{n}\right)\left(\lambda_{s} s_{i}+\lambda_{b}\right)\right),\left(2 s_{j}-1\right) \operatorname{Poi}\left(\left(1-\frac{k}{n}\right)\left(\lambda_{s} s_{j+2m}+\lambda_{b}\right)\right)\right\}\right)\mid \bm{s} \right\}  \\
		&=\mathrm{E}\left\{\left(2 s_{j}-1\right)\left(2 s_{j+2m}-1\right)\left(\left(1-\frac{k}{n}\right)\left(\lambda_{s} s_{j+2m}+\lambda_{b}\right)\right)\right\}\\
		&=\left(1-\frac{k}{n}\right)\mathrm{E}\left(4\lambda_{s}s_{j}s_{j+2m}^2-2\lambda_{s}s_{j}s_{j+2m}-2\lambda_{s}s_{j+2m}^2+\lambda_{s}s_{j+2m}\right)\\
		&+\left(1-\frac{k}{n}\right)\mathrm{E}\left(4\lambda_{b}s_{j}s_{j+2m}-2\lambda_{b}s_{j}-2\lambda_{b}s_{j+2m}+\lambda_{b}\right)\\
		&=0,
	\end{aligned}
\end{equation}

We have $ s_{j+2m} = 0$ as well as $ s_{j+2m+1} = 0$ when $(L-2m+1) \leq j \leq L $.

For $ 2m $ pair of $ \left(i,j\right) $ with $ i-1=j+2m $, 
\begin{equation}
	\begin{aligned}
		&\mathrm{E}\left(\operatorname{Cov}\left\{\left(2 s_{i}-1\right) \operatorname{Poi}\left(\varepsilon\left(\lambda_{s} s_{i-1}+\lambda_{b}\right)\right),\left(2 s_{j}-1\right) \operatorname{Poi}\left(\varepsilon\left(\lambda_{s} s_{j+2m}+\lambda_{b}\right)\right)\right\}\right)\\
		&=\mathrm{E}\left\{\mathrm{E}\left(\operatorname{Cov}\left\{\left(2 s_{i}-1\right) \operatorname{Poi}\left(\varepsilon\left(\lambda_{s} s_{i-1}+\lambda_{b}\right)\right),\left(2 s_{j}-1\right) \operatorname{Poi}\left(\varepsilon\left(\lambda_{s} s_{j+2m}+\lambda_{b}\right)\right)\right\}\right)\mid \bm{s} \right\}  \\
		&=\mathrm{E}\left\{\left(2 s_{j}-1\right)\left(2 s_{j+2m+1}-1\right)\left(\varepsilon\left(\lambda_{s} s_{j+2m}+\lambda_{b}\right)\right)\right\}\\
		&=\varepsilon\mathrm{E}\left(4\lambda_{s}s_{j}s_{j+2m}s_{j+2m+1}-2\lambda_{s}s_{j}s_{j+2m}-2\lambda_{s}s_{j+2m}s_{j+2m+1}+\lambda_{s}s_{j+2m}\right)\\
		&+\varepsilon\mathrm{E}\left(4\lambda_{b}s_{j}s_{j+2m+1}-2\lambda_{b}s_{j}-2\lambda_{b}s_{j+2m+1}+\lambda_{b}\right)\\
		&=0,\\
	\end{aligned}
\end{equation}

\begin{equation}
	\begin{aligned}
		&\mathrm{E}\left(\operatorname{Cov}\left\{\left(2 s_{i}-1\right) \operatorname{Poi}\left(\left(\frac{k}{n}-\varepsilon\right)\left(\lambda_{s} s_{i}+\lambda_{b}\right)\right),\left(2 s_{j}-1\right) \operatorname{Poi}\left(\left(\frac{k}{n}-\varepsilon\right)\left(\lambda_{s} s_{j+2m+1}+\lambda_{b}\right)\right)\right\}\right)\\
		&=\mathrm{E}\left\{\mathrm{E}\left(\operatorname{Cov}\left\{\left(2 s_{i}-1\right) \operatorname{Poi}\left(\left(\frac{k}{n}-\varepsilon\right)\left(\lambda_{s} s_{i}+\lambda_{b}\right)\right),\left(2 s_{j}-1\right) \operatorname{Poi}\left(\left(\frac{k}{n}-\varepsilon\right)\left(\lambda_{s} s_{j+2m+1}+\lambda_{b}\right)\right)\right\}\right)\mid \bm{s} \right\}  \\
		&=\mathrm{E}\left\{\left(2 s_{j}-1\right)\left(2 s_{j+2m+1}-1\right)\left(\left(\frac{k}{n}-\varepsilon\right)\left(\lambda_{s} s_{j+2m+1}+\lambda_{b}\right)\right)\right\}\\
		&=\left(\frac{k}{n}-\varepsilon\right)\mathrm{E}\left(4\lambda_{s}s_{j}s_{j+2m+1}^2-2\lambda_{s}s_{j}s_{j+2m+1}-2\lambda_{s}s_{j+2m+1}^2+\lambda_{s}s_{j+2m+1}\right)\\
		&+\left(\frac{k}{n}-\varepsilon\right)\mathrm{E}\left(4\lambda_{b}s_{j}s_{j+2m+1}-2\lambda_{b}s_{j}
		-2\lambda_{b}s_{j+2m+1}+\lambda_{b}\right)\\
		&=0,
	\end{aligned}
\end{equation}

For $ 2m $ pair of $ \left(i,j\right) $ with $ i=j+2m $,
\begin{equation}
	\begin{aligned}
		&\mathrm{E}\left(\operatorname{Cov}\left\{\left(2 s_{i}-1\right) \operatorname{Poi}\left(\left(1-\frac{k}{n}\right)\left(\lambda_{s} s_{i}+\lambda_{b}\right)\right),\left(2 s_{j}-1\right) \operatorname{Poi}\left(\left(1-\frac{k}{n}\right)\left(\lambda_{s} s_{j+2m}+\lambda_{b}\right)\right)\right\}\right)\\
		&=\mathrm{E}\left\{\mathrm{E}\left(\operatorname{Cov}\left\{\left(2 s_{i}-1\right) \operatorname{Poi}\left(\left(1-\frac{k}{n}\right)\left(\lambda_{s} s_{i}+\lambda_{b}\right)\right),\left(2 s_{j}-1\right) \operatorname{Poi}\left(\left(1-\frac{k}{n}\right)\left(\lambda_{s} s_{j+2m}+\lambda_{b}\right)\right)\right\}\right)\mid \bm{s} \right\}  \\
		&=\mathrm{E}\left\{\left(2 s_{j}-1\right)\left(2 s_{j+2m}-1\right)\left(\left(1-\frac{k}{n}\right)\left(\lambda_{s} s_{j+2m}+\lambda_{b}\right)\right)\right\}\\
		&=\left(1-\frac{k}{n}\right)\mathrm{E}\left(4\lambda_{s}s_{j}s_{j+2m}^2-2\lambda_{s}s_{j}s_{j+2m}-2\lambda_{s}s_{j+2m}^2+\lambda_{s}s_{j+2m}\right)\\
		&+\left(1-\frac{k}{n}\right)\mathrm{E}\left(4\lambda_{b}s_{j}s_{j+2m}-2\lambda_{b}s_{j}-2\lambda_{b}s_{j+2m}+\lambda_{b}\right)\\
		&=0,
	\end{aligned}
\end{equation}

On the other hand, $ s_j$ lies in $\{1,0,...\} $ part. We have $ \mathrm{E}\left(s_{j}s_{j+2m}\right) = 1$ and $ \mathrm{E}\left(s_{j}s_{j+2m+1}\right) = 0$ when $ \left(\frac{1-\alpha}{2} L +2\right) \leq j \leq \left(\frac{1+\alpha}{2}L-2m-1\right) $ with $ s_j $ not lying in boundaries


For $ (\alpha L-2m-2) $ pairs of $ \left(i,j\right) $ with $ i-1=j+2m $, 
\begin{equation}
	\begin{aligned}
		&\mathrm{E}\left(\operatorname{Cov}\left\{\left(2 s_{i}-1\right) \operatorname{Poi}\left(\varepsilon\left(\lambda_{s} s_{i-1}+\lambda_{b}\right)\right),\left(2 s_{j}-1\right) \operatorname{Poi}\left(\varepsilon\left(\lambda_{s} s_{j+2m}+\lambda_{b}\right)\right)\right\}\right)\\
		&=\mathrm{E}\left\{\mathrm{E}\left(\operatorname{Cov}\left\{\left(2 s_{i}-1\right) \operatorname{Poi}\left(\varepsilon\left(\lambda_{s} s_{i-1}+\lambda_{b}\right)\right),\left(2 s_{j}-1\right) \operatorname{Poi}\left(\varepsilon\left(\lambda_{s} s_{j+2m}+\lambda_{b}\right)\right)\right\}\right)\mid \bm{s} \right\}  \\
		&=\mathrm{E}\left\{\left(2 s_{j}-1\right)\left(2 s_{j+2m+1}-1\right)\left(\varepsilon\left(\lambda_{s} s_{j+2m}+\lambda_{b}\right)\right)\right\}\\
		&=\left(- \varepsilon \right)\left(\frac{\lambda_{s}}{2}+\lambda_{b}\right),
	\end{aligned}
\end{equation}

\begin{equation}
	\begin{aligned}
		&\mathrm{E}\left(\operatorname{Cov}\left\{\left(2 s_{i}-1\right) \operatorname{Poi}\left(\left(\frac{k}{n}-\varepsilon\right)\left(\lambda_{s} s_{i}+\lambda_{b}\right)\right),\left(2 s_{j}-1\right) \operatorname{Poi}\left(\left(\frac{k}{n}-\varepsilon\right)\left(\lambda_{s} s_{j+2m+1}+\lambda_{b}\right)\right)\right\}\right)\\
		&=\mathrm{E}\left\{\mathrm{E}\left(\operatorname{Cov}\left\{\left(2 s_{i}-1\right) \operatorname{Poi}\left(\left(\frac{k}{n}-\varepsilon\right)\left(\lambda_{s} s_{i}+\lambda_{b}\right)\right),\left(2 s_{j}-1\right) \operatorname{Poi}\left(\left(\frac{k}{n}-\varepsilon\right)\left(\lambda_{s} s_{j+2m+1}+\lambda_{b}\right)\right)\right\}\right)\mid \bm{s} \right\}  \\
		&=\mathrm{E}\left\{\left(2 s_{j}-1\right)\left(2 s_{j+2m+1}-1\right)\left(\left(\frac{k}{n}-\varepsilon\right)\left(\lambda_{s} s_{j+2m+1}+\lambda_{b}\right)\right)\right\}\\
		&= -\left(\frac{k}{n}-\varepsilon \right)\left(\frac{\lambda_{s}}{2}+\lambda_{b}\right),
	\end{aligned}
\end{equation}




For $ (\alpha L-2m-2) $ pairs of $ \left(i,j\right) $ with $ i=j+2m $,
\begin{equation}
	\begin{aligned}
		&\mathrm{E}\left(\operatorname{Cov}\left\{\left(2 s_{i}-1\right) \operatorname{Poi}\left(\left(1-\frac{k}{n}\right)\left(\lambda_{s} s_{i}+\lambda_{b}\right)\right),\left(2 s_{j}-1\right) \operatorname{Poi}\left(\left(1-\frac{k}{n}\right)\left(\lambda_{s} s_{j+2m}+\lambda_{b}\right)\right)\right\}\right)\\
		&=\mathrm{E}\left\{\mathrm{E}\left(\operatorname{Cov}\left\{\left(2 s_{i}-1\right) \operatorname{Poi}\left(\left(1-\frac{k}{n}\right)\left(\lambda_{s} s_{i}+\lambda_{b}\right)\right),\left(2 s_{j}-1\right) \operatorname{Poi}\left(\left(1-\frac{k}{n}\right)\left(\lambda_{s} s_{j+2m}+\lambda_{b}\right)\right)\right\}\right)\mid \bm{s} \right\}  \\
		&=\mathrm{E}\left\{\left(2 s_{j}-1\right)\left(2 s_{j+2m}-1\right)\left(\left(1-\frac{k}{n}\right)\left(\lambda_{s} s_{j+2m}+\lambda_{b}\right)\right)\right\}\\
		&=\left(1-\frac{k}{n}\right)\left(\frac{\lambda_{s}}{2}+\lambda_{b}\right),
	\end{aligned}
\end{equation}

If $ s_j$ lies in $\{1,0,...\} $ part, we have $ s_{j}=s_{j+2m} = 0$ and $ \mathrm{E}\left(s_{j}s_{j+2m+1}\right) = 0$ when $  j = \left(\frac{1+\alpha}{2} L-2m\right) $. $ s_{j+2m}$ is the end of $\{1,0,...\} $ part and $ s_{j+2m+1}$ lies in RANDOM part.

For $ 1 $ pair of $ \left(i,j\right) $ with $ i-1=j+2m $, 
\begin{equation}
	\begin{aligned}
		&\mathrm{E}\left(\operatorname{Cov}\left\{\left(2 s_{i}-1\right) \operatorname{Poi}\left(\varepsilon\left(\lambda_{s} s_{i-1}+\lambda_{b}\right)\right),\left(2 s_{j}-1\right) \operatorname{Poi}\left(\varepsilon\left(\lambda_{s} s_{j+2m}+\lambda_{b}\right)\right)\right\}\right)\\
		&=\mathrm{E}\left\{\mathrm{E}\left(\operatorname{Cov}\left\{\left(2 s_{i}-1\right) \operatorname{Poi}\left(\varepsilon\left(\lambda_{s} s_{i-1}+\lambda_{b}\right)\right),\left(2 s_{j}-1\right) \operatorname{Poi}\left(\varepsilon\left(\lambda_{s} s_{j+2m}+\lambda_{b}\right)\right)\right\}\right)\mid \bm{s} \right\}  \\
		&=\mathrm{E}\left\{\left(2 s_{j}-1\right)\left(2 s_{j+2m+1}-1\right)\left(\varepsilon\left(\lambda_{s} s_{j+2m}+\lambda_{b}\right)\right)\right\}\\
		&=\varepsilon\mathrm{E}\left(4\lambda_{s}s_{j}s_{j+2m}s_{j+2m+1}-2\lambda_{s}s_{j}s_{j+2m}-2\lambda_{s}s_{j+2m}s_{j+2m+1}+\lambda_{s}s_{j+2m}\right)\\
		&+\varepsilon\mathrm{E}\left(4\lambda_{b}s_{j}s_{j+2m+1}-2\lambda_{b}s_{j}-2\lambda_{b}s_{j+2m+1}+\lambda_{b}\right)\\
		&=0,\\
	\end{aligned}
\end{equation}

\begin{equation}
	\begin{aligned}
		&\mathrm{E}\left(\operatorname{Cov}\left\{\left(2 s_{i}-1\right) \operatorname{Poi}\left(\left(\frac{k}{n}-\varepsilon\right)\left(\lambda_{s} s_{i}+\lambda_{b}\right)\right),\left(2 s_{j}-1\right) \operatorname{Poi}\left(\left(\frac{k}{n}-\varepsilon\right)\left(\lambda_{s} s_{j+2m+1}+\lambda_{b}\right)\right)\right\}\right)\\
		&=\mathrm{E}\left\{\mathrm{E}\left(\operatorname{Cov}\left\{\left(2 s_{i}-1\right) \operatorname{Poi}\left(\left(\frac{k}{n}-\varepsilon\right)\left(\lambda_{s} s_{i}+\lambda_{b}\right)\right),\left(2 s_{j}-1\right) \operatorname{Poi}\left(\left(\frac{k}{n}-\varepsilon\right)\left(\lambda_{s} s_{j+2m+1}+\lambda_{b}\right)\right)\right\}\right)\mid \bm{s} \right\}  \\
		&=\mathrm{E}\left\{\left(2 s_{j}-1\right)\left(2 s_{j+2m+1}-1\right)\left(\left(\frac{k}{n}-\varepsilon\right)\left(\lambda_{s} s_{j+2m+1}+\lambda_{b}\right)\right)\right\}\\
		&=\left(\frac{k}{n}-\varepsilon\right)\mathrm{E}\left(4\lambda_{s}s_{j}s_{j+2m+1}^2-2\lambda_{s}s_{j}s_{j+2m+1}-2\lambda_{s}s_{j+2m+1}^2+\lambda_{s}s_{j+2m+1}\right)\\
		&+\left(\frac{k}{n}-\varepsilon\right)\mathrm{E}\left(4\lambda_{b}s_{j}s_{j+2m+1}-2\lambda_{b}s_{j}
		-2\lambda_{b}s_{j+2m+1}+\lambda_{b}\right)\\
		&=-\left(\frac{k}{n}-\varepsilon\right)\frac{\lambda_{s}}{2},
	\end{aligned}
\end{equation}

For $ 1 $ pair of $ \left(i,j\right) $ with $ i=j+2m $,
\begin{equation}
	\begin{aligned}
		&\mathrm{E}\left(\operatorname{Cov}\left\{\left(2 s_{i}-1\right) \operatorname{Poi}\left(\left(1-\frac{k}{n}\right)\left(\lambda_{s} s_{i}+\lambda_{b}\right)\right),\left(2 s_{j}-1\right) \operatorname{Poi}\left(\left(1-\frac{k}{n}\right)\left(\lambda_{s} s_{j+2m}+\lambda_{b}\right)\right)\right\}\right)\\
		&=\mathrm{E}\left\{\mathrm{E}\left(\operatorname{Cov}\left\{\left(2 s_{i}-1\right) \operatorname{Poi}\left(\left(1-\frac{k}{n}\right)\left(\lambda_{s} s_{i}+\lambda_{b}\right)\right),\left(2 s_{j}-1\right) \operatorname{Poi}\left(\left(1-\frac{k}{n}\right)\left(\lambda_{s} s_{j+2m}+\lambda_{b}\right)\right)\right\}\right)\mid \bm{s} \right\}  \\
		&=\mathrm{E}\left\{\left(2 s_{j}-1\right)\left(2 s_{j+2m}-1\right)\left(\left(1-\frac{k}{n}\right)\left(\lambda_{s} s_{j+2m}+\lambda_{b}\right)\right)\right\}\\
		&=\left(1-\frac{k}{n}\right)\mathrm{E}\left(4\lambda_{s}s_{j}s_{j+2m}^2-2\lambda_{s}s_{j}s_{j+2m}-2\lambda_{s}s_{j+2m}^2+\lambda_{s}s_{j+2m}\right)\\
		&+\left(1-\frac{k}{n}\right)\mathrm{E}\left(4\lambda_{b}s_{j}s_{j+2m}-2\lambda_{b}s_{j}-2\lambda_{b}s_{j+2m}+\lambda_{b}\right)\\
		&=\left(1-\frac{k}{n}\right)\lambda_{b},
	\end{aligned}
\end{equation}

Paired with the boundary above, we have $ s_{j}=s_{j+2m}=1$ and $ \mathrm{E}\left(s_{j}s_{j+2m+1}\right) = 0$ when $  j = \left(\frac{1-\alpha}{2} L +1\right) $. $ s_{j} $ is the start of $\{1,0,...\} $ part.

For $ 1 $ pair of $ \left(i,j\right) $ with $ i-1=j+2m $, 
\begin{equation}
	\begin{aligned}
		&\mathrm{E}\left(\operatorname{Cov}\left\{\left(2 s_{i}-1\right) \operatorname{Poi}\left(\varepsilon\left(\lambda_{s} s_{i-1}+\lambda_{b}\right)\right),\left(2 s_{j}-1\right) \operatorname{Poi}\left(\varepsilon\left(\lambda_{s} s_{j+2m}+\lambda_{b}\right)\right)\right\}\right)\\
		&=\mathrm{E}\left\{\mathrm{E}\left(\operatorname{Cov}\left\{\left(2 s_{i}-1\right) \operatorname{Poi}\left(\varepsilon\left(\lambda_{s} s_{i-1}+\lambda_{b}\right)\right),\left(2 s_{j}-1\right) \operatorname{Poi}\left(\varepsilon\left(\lambda_{s} s_{j+2m}+\lambda_{b}\right)\right)\right\}\right)\mid \bm{s} \right\}  \\
		&=\mathrm{E}\left\{\left(2 s_{j}-1\right)\left(2 s_{j+2m+1}-1\right)\left(\varepsilon\left(\lambda_{s} s_{j+2m}+\lambda_{b}\right)\right)\right\}\\
		&=\varepsilon\mathrm{E}\left(4\lambda_{s}s_{j}s_{j+2m}s_{j+2m+1}-2\lambda_{s}s_{j}s_{j+2m}-2\lambda_{s}s_{j+2m}s_{j+2m+1}+\lambda_{s}s_{j+2m}\right)\\
		&+\varepsilon\mathrm{E}\left(4\lambda_{b}s_{j}s_{j+2m+1}-2\lambda_{b}s_{j}-2\lambda_{b}s_{j+2m+1}+\lambda_{b}\right)\\
		&=-\varepsilon\left(\lambda_{s}+\lambda_{b}\right),\\
	\end{aligned}
\end{equation}

\begin{equation}
	\begin{aligned}
		&\mathrm{E}\left(\operatorname{Cov}\left\{\left(2 s_{i}-1\right) \operatorname{Poi}\left(\left(\frac{k}{n}-\varepsilon\right)\left(\lambda_{s} s_{i}+\lambda_{b}\right)\right),\left(2 s_{j}-1\right) \operatorname{Poi}\left(\left(\frac{k}{n}-\varepsilon\right)\left(\lambda_{s} s_{j+2m+1}+\lambda_{b}\right)\right)\right\}\right)\\
		&=\mathrm{E}\left\{\mathrm{E}\left(\operatorname{Cov}\left\{\left(2 s_{i}-1\right) \operatorname{Poi}\left(\left(\frac{k}{n}-\varepsilon\right)\left(\lambda_{s} s_{i}+\lambda_{b}\right)\right),\left(2 s_{j}-1\right) \operatorname{Poi}\left(\left(\frac{k}{n}-\varepsilon\right)\left(\lambda_{s} s_{j+2m+1}+\lambda_{b}\right)\right)\right\}\right)\mid \bm{s} \right\}  \\
		&=\mathrm{E}\left\{\left(2 s_{j}-1\right)\left(2 s_{j+2m+1}-1\right)\left(\left(\frac{k}{n}-\varepsilon\right)\left(\lambda_{s} s_{j+2m+1}+\lambda_{b}\right)\right)\right\}\\
		&=\left(\frac{k}{n}-\varepsilon\right)\mathrm{E}\left(4\lambda_{s}s_{j}s_{j+2m+1}^2-2\lambda_{s}s_{j}s_{j+2m+1}-2\lambda_{s}s_{j+2m+1}^2+\lambda_{s}s_{j+2m+1}\right)\\
		&+\left(\frac{k}{n}-\varepsilon\right)\mathrm{E}\left(4\lambda_{b}s_{j}s_{j+2m+1}-2\lambda_{b}s_{j}
		-2\lambda_{b}s_{j+2m+1}+\lambda_{b}\right)\\
		&=-\left(\frac{k}{n}-\varepsilon\right)\lambda_{b},
	\end{aligned}
\end{equation}

For $ 1 $ pair of $ \left(i,j\right) $ with $ i=j+2m $,
\begin{equation}
	\begin{aligned}
		&\mathrm{E}\left(\operatorname{Cov}\left\{\left(2 s_{i}-1\right) \operatorname{Poi}\left(\left(1-\frac{k}{n}\right)\left(\lambda_{s} s_{i}+\lambda_{b}\right)\right),\left(2 s_{j}-1\right) \operatorname{Poi}\left(\left(1-\frac{k}{n}\right)\left(\lambda_{s} s_{j+2m}+\lambda_{b}\right)\right)\right\}\right)\\
		&=\mathrm{E}\left\{\mathrm{E}\left(\operatorname{Cov}\left\{\left(2 s_{i}-1\right) \operatorname{Poi}\left(\left(1-\frac{k}{n}\right)\left(\lambda_{s} s_{i}+\lambda_{b}\right)\right),\left(2 s_{j}-1\right) \operatorname{Poi}\left(\left(1-\frac{k}{n}\right)\left(\lambda_{s} s_{j+2m}+\lambda_{b}\right)\right)\right\}\right)\mid \bm{s} \right\}  \\
		&=\mathrm{E}\left\{\left(2 s_{j}-1\right)\left(2 s_{j+2m}-1\right)\left(\left(1-\frac{k}{n}\right)\left(\lambda_{s} s_{j+2m}+\lambda_{b}\right)\right)\right\}\\
		&=\left(1-\frac{k}{n}\right)\mathrm{E}\left(4\lambda_{s}s_{j}s_{j+2m}^2-2\lambda_{s}s_{j}s_{j+2m}-2\lambda_{s}s_{j+2m}^2+\lambda_{s}s_{j+2m}\right)\\
		&+\left(1-\frac{k}{n}\right)\mathrm{E}\left(4\lambda_{b}s_{j}s_{j+2m}-2\lambda_{b}s_{j}-2\lambda_{b}s_{j+2m}+\lambda_{b}\right)\\
		&=\left(1-\frac{k}{n}\right)(\lambda_{s}+\lambda_{b}),
	\end{aligned}
\end{equation}

If $ s_j$ lies in $\{1,0,...\} $ part, we have $ \mathrm{E}\left(s_{j}s_{j+2m}\right) = \frac{1}{4}$ and $ \mathrm{E}\left(s_{j}s_{j+2m+1}\right) = \frac{1}{4}$ when $ \left(\frac{1+\alpha}{2} L-2m+1 \leq j \leq \frac{1+\alpha}{2} L\right) $. $ s_{j+2m} $ and $ s_{j+2m+1} $ lie in RANDOM part.

For $ 2m $ pair of $ \left(i,j\right) $ with $ i-1=j+2m $, 
\begin{equation}
	\begin{aligned}
		&\mathrm{E}\left(\operatorname{Cov}\left\{\left(2 s_{i}-1\right) \operatorname{Poi}\left(\varepsilon\left(\lambda_{s} s_{i-1}+\lambda_{b}\right)\right),\left(2 s_{j}-1\right) \operatorname{Poi}\left(\varepsilon\left(\lambda_{s} s_{j+2m}+\lambda_{b}\right)\right)\right\}\right)\\
		&=\mathrm{E}\left\{\mathrm{E}\left(\operatorname{Cov}\left\{\left(2 s_{i}-1\right) \operatorname{Poi}\left(\varepsilon\left(\lambda_{s} s_{i-1}+\lambda_{b}\right)\right),\left(2 s_{j}-1\right) \operatorname{Poi}\left(\varepsilon\left(\lambda_{s} s_{j+2m}+\lambda_{b}\right)\right)\right\}\right)\mid \bm{s} \right\}  \\
		&=\mathrm{E}\left\{\left(2 s_{j}-1\right)\left(2 s_{j+2m+1}-1\right)\left(\varepsilon\left(\lambda_{s} s_{j+2m}+\lambda_{b}\right)\right)\right\}\\
		&=\varepsilon\mathrm{E}\left(4\lambda_{s}s_{j}s_{j+2m}s_{j+2m+1}-2\lambda_{s}s_{j}s_{j+2m}-2\lambda_{s}s_{j+2m}s_{j+2m+1}+\lambda_{s}s_{j+2m}\right)\\
		&+\varepsilon\mathrm{E}\left(4\lambda_{b}s_{j}s_{j+2m+1}-2\lambda_{b}s_{j}-2\lambda_{b}s_{j+2m+1}+\lambda_{b}\right)\\
		&=0,\\
	\end{aligned}
\end{equation}

\begin{equation}
	\begin{aligned}
		&\mathrm{E}\left(\operatorname{Cov}\left\{\left(2 s_{i}-1\right) \operatorname{Poi}\left(\left(\frac{k}{n}-\varepsilon\right)\left(\lambda_{s} s_{i}+\lambda_{b}\right)\right),\left(2 s_{j}-1\right) \operatorname{Poi}\left(\left(\frac{k}{n}-\varepsilon\right)\left(\lambda_{s} s_{j+2m+1}+\lambda_{b}\right)\right)\right\}\right)\\
		&=\mathrm{E}\left\{\mathrm{E}\left(\operatorname{Cov}\left\{\left(2 s_{i}-1\right) \operatorname{Poi}\left(\left(\frac{k}{n}-\varepsilon\right)\left(\lambda_{s} s_{i}+\lambda_{b}\right)\right),\left(2 s_{j}-1\right) \operatorname{Poi}\left(\left(\frac{k}{n}-\varepsilon\right)\left(\lambda_{s} s_{j+2m+1}+\lambda_{b}\right)\right)\right\}\right)\mid \bm{s} \right\}  \\
		&=\mathrm{E}\left\{\left(2 s_{j}-1\right)\left(2 s_{j+2m+1}-1\right)\left(\left(\frac{k}{n}-\varepsilon\right)\left(\lambda_{s} s_{j+2m+1}+\lambda_{b}\right)\right)\right\}\\
		&=\left(\frac{k}{n}-\varepsilon\right)\mathrm{E}\left(4\lambda_{s}s_{j}s_{j+2m+1}^2-2\lambda_{s}s_{j}s_{j+2m+1}-2\lambda_{s}s_{j+2m+1}^2+\lambda_{s}s_{j+2m+1}\right)\\
		&+\left(\frac{k}{n}-\varepsilon\right)\mathrm{E}\left(4\lambda_{b}s_{j}s_{j+2m+1}-2\lambda_{b}s_{j}
		-2\lambda_{b}s_{j+2m+1}+\lambda_{b}\right)\\
		&=0,
	\end{aligned}
\end{equation}

For $ 2m $ pair of $ \left(i,j\right) $ with $ i=j+2m $,
\begin{equation}
	\begin{aligned}
		&\mathrm{E}\left(\operatorname{Cov}\left\{\left(2 s_{i}-1\right) \operatorname{Poi}\left(\left(1-\frac{k}{n}\right)\left(\lambda_{s} s_{i}+\lambda_{b}\right)\right),\left(2 s_{j}-1\right) \operatorname{Poi}\left(\left(1-\frac{k}{n}\right)\left(\lambda_{s} s_{j+2m}+\lambda_{b}\right)\right)\right\}\right)\\
		&=\mathrm{E}\left\{\mathrm{E}\left(\operatorname{Cov}\left\{\left(2 s_{i}-1\right) \operatorname{Poi}\left(\left(1-\frac{k}{n}\right)\left(\lambda_{s} s_{i}+\lambda_{b}\right)\right),\left(2 s_{j}-1\right) \operatorname{Poi}\left(\left(1-\frac{k}{n}\right)\left(\lambda_{s} s_{j+2m}+\lambda_{b}\right)\right)\right\}\right)\mid \bm{s} \right\}  \\
		&=\mathrm{E}\left\{\left(2 s_{j}-1\right)\left(2 s_{j+2m}-1\right)\left(\left(1-\frac{k}{n}\right)\left(\lambda_{s} s_{j+2m}+\lambda_{b}\right)\right)\right\}\\
		&=\left(1-\frac{k}{n}\right)\mathrm{E}\left(4\lambda_{s}s_{j}s_{j+2m}^2-2\lambda_{s}s_{j}s_{j+2m}-2\lambda_{s}s_{j+2m}^2+\lambda_{s}s_{j+2m}\right)\\
		&+\left(1-\frac{k}{n}\right)\mathrm{E}\left(4\lambda_{b}s_{j}s_{j+2m}-2\lambda_{b}s_{j}-2\lambda_{b}s_{j+2m}+\lambda_{b}\right)\\
		&=0,
	\end{aligned}
\end{equation}

By adding up components above, the mathematical expectation of covariance is,
\begin{align}
	\mathrm{E}\left\{\operatorname{Cov}\left(\operatorname{C}_{0,0}, \operatorname{C}_{2m, k}\right)\right\}&=(\alpha L-2 m-2)\left(\frac{\lambda_{s}}{2}+\lambda_{b}\right)\left(1-2 \frac{k}{n}\right)-\left(\frac{k}{n}-\varepsilon\right)\frac{\lambda_{s}}{2}+\left(1-\frac{k}{n}\right)\lambda_{b}\\
	&-\varepsilon\left(\lambda_{s}+\lambda_{b}\right)-\left(\frac{k}{n}-\varepsilon\right)\lambda_{b}+\left(1-\frac{k}{n}\right)(\lambda_{s}+\lambda_{b})\\
	&=\left(\frac{\lambda_{s}}{2}+\lambda_{b}\right)\left[\left(\alpha L -2m \right)\left(1-2\frac{k}{n}\right)\right]+\frac{k}{n}\left(\frac{\lambda_{s}}{2}+\lambda_{b}\right)-\varepsilon\frac{\lambda_{s}}{2}.
\end{align}

\end{appendices}

\begin{footnotesize}
	\bibliographystyle{IEEEtran}
	\bibliography{Synchronization_Sequence_ShihuiYu}
\end{footnotesize}

\end{document}